\documentclass[submission,copyright,creativecommons,fleqn]{eptcs}

\providecommand{\event}{EXPRESS/SOS 2017} 

\usepackage{breakurl}             
\usepackage{hyperref}
\hypersetup{bookmarks=true,bookmarksnumbered=true,bookmarksopen}
\usepackage{amsmath}
\usepackage{amsthm}
\usepackage[usenames]{color}
\usepackage{amssymb}
\usepackage{procalg}
\usepackage{tikz}
\usetikzlibrary{arrows,automata}
\usepackage{amscd}
\usepackage{txfonts}
\usepackage{amsfonts,semantic,colortbl,mathrsfs,stmaryrd,mathtools}
\usepackage{epsfig,graphicx,subfigure}
\usepackage{enumerate,enumitem}
\usepackage{multirow,multicol}
\usepackage{tabularx}
\usepackage{gastex}

\usepackage{color}
\makeatletter
\@ifundefined{ifAP}{%

\newcommand{\todo}[1]{\underline{\sf todo}: {\color{red}{#1}}}
}{
\ifAP
\else
\fi
\newcommand{\todo}[1]{}
}
\makeatother

\newenvironment{tarray}[2][c]{
  \settowidth{\dimen1}{${} = {}$}%
  \setlength{\arraycolsep}{0.125\dimen1}%
  \hspace{-0.125\dimen1}\array[#1]{#2}\relax
}
{
   \endarray\hspace{-0.125\dimen1}
}
\def\tbb{\begin{tarray}{llll}}
\def\tbbt{\begin{tarray}[t]{llll}}
\def\tee{\end{tarray}}

\renewcommand{\bullet}{\ensuremath{;}}
\newcommand{\myparagraph}[1]{\noindent\textbf{#1}}
\newcommand{\T}{\ensuremath{\mathcal{T}} }
\newcommand{\A}{\ensuremath{\mathcal{A}}}
\newcommand{\Atau}{\ensuremath{\mathcal{A}_{\tau}}}
\newcommand{\Sta}{\ensuremath{\mathcal{S}}}

\newcommand{\R}{\mathcal{\mathop{R}}}
\newcommand{\M}{\mathcal{M}}
\newcommand{\F}{\mathcal{F}}
\newcommand{\D}{\mathcal{D}}
\newcommand{\N}{\mathcal{N}}
\newcommand{\I}{\mathcal{I}}
\newcommand{\C}{\mathcal{C}}
\newcommand{\V}{\mathcal{V}}
\newcommand{\ddef}{\overset{\textrm{def}}{=}}
\newcommand{\Dbox}{\mathcal{D}_{\Box}}
\newcommand{\tphd}[1]{\check{#1}}
\newcommand{\tphdL}[1]{{#1 \!}^{\scriptscriptstyle <}}
\newcommand{\tphdR}[1]{\prescript{\scriptscriptstyle >}{}{\! #1}}
\newcommand{\Conf}[1][]{\mathcal{C}_{#1}}
\newcommand{\length}[1]{\ensuremath{\mathit{length}(#1)}}
\newcommand{\stack}[1]{\ensuremath{\mathit{stack}(#1)}}
\newcommand{\get}[2]{\ensuremath{\mathit{get}(#1,#2)}}
\newcommand{\suffset}[2]{\ensuremath{\mathit{suffset}(#1,#2)}}
\newcommand{\ar}{\ensuremath{\mathit{ar}}}
\newcommand{\bbisimd}{\ensuremath{\mathrel{\bbisim^{\Delta}}}}
\newcommand{\rbbisimd}{\ensuremath{\mathrel{\rbbisim^{\Delta}}}}
\newcommand{\nil}{\ensuremath{\mathalpha{\mathbf{0}}}}
\newcommand{\one}{\ensuremath{\mathalpha{\mathbf{1}}}}
\newcommand{\NT}[1]{\ensuremath{\mathalpha{NT}(#1)}}
\newcommand{\encode}[1]{\lceil #1\rceil}

\newcommand{\nesting}{\ensuremath{\mathalpha{\sharp}}}
\newcommand{\pushdown}{\ensuremath{\mathalpha{\$}}}
\newcommand{\backforth}{\ensuremath{\mathalpha{\leftrightarrows}}}
\newcommand{\replication}{\ensuremath{\mathalpha{!}}}
\newcommand{\TCP}{\ensuremath{\textrm{TCP}}}
\newcommand{\TCPN}{\ensuremath{\textrm{TCP}^{\nesting}}}
\newcommand{\TCPP}{\ensuremath{\textrm{TCP}^{\pushdown}}}
\newcommand{\TCPB}{\ensuremath{\textrm{TCP}^{\backforth}}}
\newcommand{\TCPR}{\ensuremath{\textrm{TCP}^{\replication}}}
\newcommand{\TCPS}{\ensuremath{\textrm{TCP}^{\bullet}}}
\newcommand{\CCS}{\ensuremath{\textrm{CCS}}}
\newcommand{\SA}{\ensuremath{\textrm{TSP}}}
\newcommand{\SAS}{\ensuremath{\textrm{TSP}^{\bullet}}}
\newcommand{\CCSR}{\ensuremath{\textrm{CCS}^{\replication}}}
\newcommand{\PDA}{\ensuremath{\textrm{PDA}}}

\newcommand{\pushd}[2]{\ensuremath{\mathalpha{{#1}^{\pushdown}{#2}}}}
\newcommand{\backf}[2]{\ensuremath{\mathalpha{{#1}^{\backforth}{#2}}}}
\newcommand{\nest}[2]{\ensuremath{\mathalpha{{#1}^{\nesting}{#2}}}}
\newcommand{\delete}[1]{}

\newcommand{\conf}[1]{}
\newcommand{\arx}[1]{#1}

\newtheorem{definition}{Definition}
\newtheorem{lemma}{Lemma}

\newtheorem{theorem}{Theorem}

\title{Sequential Composition in the Presence of Intermediate Termination}
\author{Jos Baeten
\institute{CWI\\ Amsterdam, the Netherlands}
\institute{University of Amsterdam,\\ Amsterdam, the Netherlands}
\email{Jos.Baeten@cwi.nl}
\and
Bas Luttik
\institute{Eindhoven University of Technology\\ Eindhoven, The Netherlands}
\email{s.p.luttik@tue.nl}
\and
Fei Yang
\institute{Eindhoven University of Technology\\ Eindhoven, The Netherlands}
\email{\quad f.yang@tue.nl}
}
\def\titlerunning{Sequential Composition in the Presence of Intermediate Termination}
\def\authorrunning{Jos Baeten, Bas Luttik \& Fei Yang}

\begin{document}
\maketitle

\begin{abstract}
The standard operational semantics of the sequential composition operator gives rise to unbounded branching and forgetfulness when transparent process expressions are put in sequence. Due to transparency, the correspondence between context-free and push-down processes fails modulo bisimilarity, and it is not clear how to specify an always terminating half counter. We propose a revised operational semantics for the sequential composition operator in the context of intermediate termination.  With the revised operational semantics, we eliminate transparency. As a consequence, we establish a correspondence between context-free processes and pushdown processes. Moreover, we prove the reactive Turing powerfulness of TCP with iteration and nesting with the revised operational semantics for sequential composition.
\end{abstract}

\section{Introduction}\label{sec:introduction}
\delete{\todo{\begin{enumerate}
\item We need to illustrate the origin of the old version of sequential composition operator.
\item We shall illustrate the role of termination.
\item We shall illustrate the problems of the old version: transparency.
\item We shall introduce the problems of pushdown automata and context free process and reactive Turing powerful process calculus and our results.
\end{enumerate}}}
The integration of the concurrency theory and the classical theory of formal languages has been extensively studied in recent years~\cite{BCLT2009}. A lot of notions in the classical theory have their counterparts in the concurrency theory~\cite{vanTilburg2011}. However, we still cannot conclude a complete correspondence for all the notions. As far as we are concerned, a major obstacle is the phenomenon of transparency of the sequential composition operator in the presence of intermediate termination.

Sequential composition is a standard operator in many process calculi. The functionality of the sequential composition operator is to concatenate the behaviours of two systems. It has been widely used in many process calculi with the notation ``$\cdot$''. We illustrate its operational semantics by a process $P\cdot Q$ in \TCP~\cite{BBR2010}. If the process $P$ has a transition $P\step{a}P'$ for some action label $a$, then the composition $P\cdot Q$ has the transition $P\cdot Q\step{a}P'\cdot Q$.
Termination is an important behaviour for models of computation~\cite{BBR2010}.\delete{ In \TCP, we use ``$\downarrow$'' to represent that a state is terminating.} Combining with sequential composition, we have two additional rules in the operational semantics. The first one states that $P\cdot Q$ terminates if both $P$ and $Q$ terminates; and the second one states that if $P$ terminates, and there is a transition $Q\step{a}Q'$ for some action label $a$, then we have the transition $P\cdot Q\step{a}Q'$. Together with the first rules, we are able to characterise the behaviour of systems consisting of two concatenated parts. The system may make a transition of the first part. If the first part is terminating, then the system may choose to skip the first part and make a transition of the second part. If both parts are terminating, then the combined system is also terminating.

In this paper, we discuss a complication on the standard version of the operational semantics of the sequential composition operator. The complication is on the transparency caused by sequential composition operator and termination~\cite{BCLT2009}. A process expression is called transparent if it is terminating. Actually, it is then transparent in a ``sequential context''. We have observed two disadvantages on transparency in the following research problems.

The relationship between context-free processes and pushdown automaton has been extensively discussed in the literature~\cite{BCvT2008}. It has been shown that every context-free process can be specified by a pushdown process modulo contra simulation. However, such result is no longer valid modulo rooted branching bisimulation, which is a finer behavioural equivalence. By stacking unboundedly many transparent terms with sequential composition, we would get a transition system with unboundedly branching on its behaviour. It was shown that such unboundedly branching behaviour cannot be specified by any pushdown process modulo rooted branching bisimulation~\cite{BCvT2008}. In order to improve the result to a finer notion of behaviour equivalence, we need to eliminate the problem of unboundedly branching.

Another problem is that transparency makes a stack of transparent process expressions forgetful. A notion of reactive Turing machines (RTM)~\cite{BLT2013} was introduced as a model to integrate concurrency and computability. The transition systems associated with RTMs are called executable. We use the RTM as a criteria of absolute expressiveness of process calculi and interactive computation models~\cite{LY15,LY16}. A process calculus is called reactively Turing powerful if every executable transition system can be specified. The process calculus \TCP{} with iteration and nesting is Turing complete~\cite{Bergstra1994,bergstra2001non}. Moreover, it follows from the result in~\cite{bergstra2001non} that it is reactively Turing powerful if termination is out of consideration. However, it is not clear to us how to reconstruct the proof of reactive Turing powerfulness if termination is considered. Due to the forgetfulness on the stacking of transparent process expressions, it is not clear to us how to define a counter that is always terminating which is crucial to establish the reactive Turing powerfulness.

In order to avoid the unwanted feature of unbounded branching and forgetfulness, we propose a revised operational semantics for the sequential composition operator. Intuitively, we disallow the transition from the second component of a sequential composition if the first component is able to still perform a transition. Thus, we avoid the problems mentioned above with the revised operator. We shall prove that every context-free process is bisimilar to a pushdown process and \TCP{} with iteration and nesting is reactively Turing powerful modulo divergence-preserving branching bisimilarity without using recursive specification in the revised semantics.

The paper is structured as follows. We first introduce TCP with the standard version of sequential composition in Section~\ref{sec:preliminaries}. Next, we discuss the complications caused by transparency in Section~\ref{sec:transparency}. Then, in Section~\ref{sec:sequential}, we propose the revised operational semantics of the sequential composition operator, and show that rooted divergence-preserving branching bisimulation is a congruence. In Section~\ref{sec:cfg}, we revisit the problem on the relationship between context-free processes and pushdown automaton, and show that every context-free process is bisimilar to a pushdown process in our revised semantics. In Section~\ref{sec:Termination}, we prove that \TCP{} with iteration and nesting is a reactively Turing powerful in the revised semantics. In Section~\ref{sec:conclusion}, we draw some conclusions and propose some future work.
\conf{The detailed proofs of our results are provided in an appendix. We also put a full version on the arxiv.}

\section{Preliminaries}\label{sec:preliminaries}
We start with introducing a notion of labelled transition systems which is used as the standard mathematical representation of behaviour. We consider transition systems with a subset of states marked with final states. We let $\A$ be a set of \emph{action symbols}, and we extend $\A$ with a special symbol $\tau\notin \A$, which intuitively denotes unobservable internal activity of the system. We shall abbreviate $\A \cup\{\tau\}$ by $\Atau$.
\begin{definition}
~\label{def:lts}
An \emph{$\Atau$-labelled transition system} is a tuple $(\Sta,\step{},\uparrow,\downarrow)$, where
\begin{enumerate}
    \item $\Sta$ is a set of \emph{states},
    \item ${\step{}}\subseteq{\Sta\times\Atau\times\Sta}$ is an $\Atau$-labelled \emph{transition relation},
    \item ${\uparrow}\in\Sta$ is the initial state, and
    \item ${\downarrow}\subseteq\Sta$ is a set of terminating states.
\end{enumerate}
\end{definition}
\delete{
\begin{definition}
A \emph{signature} $\Sigma$ consists of:
\begin{enumerate}
\item an infinite set of variables $x,y,z,\ldots$; and
\item a set $\F$ of function symbols $f,g,h,\ldots$, where each symbol $f$ has an arity $\ar(f)$.
\end{enumerate}
A function symbol of arity zero is called a \emph{constant}.
\end{definition}

\begin{definition}
Let $\Sigma$ be a signature. The collection $\mathbb{T}(\Sigma)$ of \emph{(open) terms} $p,q,r,\ldots$ over $\Sigma$ is defined as the least set satisfying:
\begin{enumerate}
\item each variable is in $\mathbb{T}(\Sigma)$;
\item if $t_1,\ldots,t_{\ar(f)}\in\mathbb{T}(\Sigma)$, then $f(t_1,\ldots,t_{\ar(f)})\in\mathbb{T}(\Sigma)$.
\end{enumerate}
A term is \emph{closed} if it does not contain any free variables. The set of closed terms is denoted by $\mathbb{C}(\Sigma)$.
\end{definition}
}

Next, we shall use the process calculus \TCP{} that allows us to describe transition systems. Let $\C$ be a set of \emph{channels} and $\Dbox$ be a set of \emph{data symbols}. For every subset $\C'\subseteq\C$, we define a special set of actions $\I_{\C'}\subseteq \Atau$ by:
\begin{equation*}
    \I_{\C'}=\{c?d,c!d\mid d\in\Dbox,c\in \C'\}
\enskip.
\end{equation*}

The actions $c?d$ and $c!d$ denote the events that a datum $d$ is received or sent along channel $c$. Furthermore, let $\N$ be a countably infinite set of names. The set of process expressions $\mathcal{P}$ is generated by the following grammar $(a\in\Atau,\,N\in\N,\,\C′\in \C)$:
\begin{equation*}
P=\nil\mid \one\mid a.P\mid P_1\cdot P_2\mid [P_1\parallel P_2]_{\C'}\mid P_1+P_2\mid N
\enskip.
\end{equation*}

We briefly comment on the operators in this syntax. The constant $\nil$ denotes \emph{deadlock}, the unsuccessfully terminated process. The constant $\one$ denotes \emph{termination}, the successfully terminated process. For each action $a\in\Atau$ there is a unary operator $a.$ denoting action prefix; the process denoted by $a.P$ can do an $a$-labelled transition to the process $P$. The binary operator $+$ denotes alternative composition or choice. The binary operator $[\_\parallel\_ ]_{\C′}$ deviates from TCP in~\cite{BBR2010} which denotes a special kind of parallel composition. It enforces communication along the channels in $\C′$, and communication results in $\tau$. The binary operator $\cdot$ represents the sequential composition of two processes.

Let $P$ be an arbitrary process expression; and we use an abbreviation inductively defined by:
\begin{enumerate}
\item $P^{0}=\one$; and
\item $P^{n+1}=P\cdot P^{n}$ for all $n\in\mathbb{N}$.
\end{enumerate}

A recursive specification $E$ is a set of equations
\begin{equation*}
E = \{N\ddef P | N\in \N, P\in\mathcal{P}\}
\enskip.
\end{equation*}
satisfying the requirements that
\begin{enumerate}
\item for every $N\in\N$ it includes at most one equation with $N$ as left-hand side,
which is referred to as the \emph{defining equation} for $N$; and
\item if some name $N'$ occurs in the right-hand side $P′$ of some equation $N′ = p′$ in
$E$, then $E$ must include a defining equation for $N'$.
\end{enumerate}
A recursive specification is \emph{guarded} if every summand in the specification that is not $\one$ is in the scope of some action prefix. 

We use structural operational semantics to associate a transition relation with process expressions defined in \TCP. We let $\step{}$ be the $\Atau$-labelled transition relation induced on the set of process expressions by operational rules in Figure~\ref{fig:semantics-tcp}. Note that we presuppose a recursive specification $E$.

\begin{figure}
\fbox{
\begin{minipage}[t]{1\textwidth}
\begin{eqnarray*}
&\inference{}{\one\downarrow}\quad\inference{\,}{a.P\step{a}P}\\
&\inference{P_1\step{a}P_1'}{P_1+ P_2\step{a}P_1'}\quad \inference{P_2\step{a}P_2'}{P_1+ P_2\step{a}P_2'}\quad
\inference{P_1\downarrow}{P_1+P_2\downarrow}\quad \inference{P_2\downarrow}{P_1+P_2\downarrow}\\
&\inference{P_1\step{a}P_1'\quad a\notin \I_{\C'}}{[P_1\parallel P_2]_{\C'}\step{a}[P_1'\parallel P_2]_{\C'}}\quad\inference{P_2\step{a}P_2'\quad a\notin\I_{\C'}}{[P_1\parallel P_2]_{\C'}\step{a}[P_1\parallel P_2']_{\C'}}\\
&\inference{P_1\step{c?d}P_1'\quad P_2\step{c!d}P_2'\quad c\in\C'}{[P_1\parallel P_2]_{\C'}\step{\tau}[P_1'\parallel P_2']_{\C'}}\quad \inference{P_1\step{c!d}P_1'\quad P_2\step{c?d}P_2'\quad c\in\C'}{[P_1\parallel P_2]_{\C'}\step{\tau}[P_1'\parallel P_2']_{\C'}}\quad \inference{P_1\downarrow\quad P_2\downarrow}{[P_1\parallel P_2]_{\C'}\downarrow}\\
& \inference{P_1\downarrow\quad  P_2\downarrow}{P_1\cdot P_2\downarrow}\quad
\inference{P_1\step{a}P_1'}{P_1\cdot P_2\step{a}P_1'\cdot P_2}\quad \inference{P_1\downarrow\quad P_2\step{a}P_2'}{P_1\cdot P_2\step{a}P_2'}\\
&\inference{P\step{a}P'\quad (N\ddef P)\in E}{N\step{a}P'}\quad\inference{P\downarrow\quad (N\ddef P)\in E}{N\downarrow}
\end{eqnarray*}
\end{minipage}
}
\caption{The operational Semantics of TCP}\label{fig:semantics-tcp}
\end{figure}
Here we use $P\step{a}P'$ to denote an $a$-labelled transition $(P,a,P')\in{\step{}}$. We say a process expression $P'$ is \emph{reachable} from $P$ is there exists process expressions $P_0,P_1,\ldots,P_n$ and labels $a_1,\ldots,a_n$, such that $P=P_0\step{a_1}P_1\ldots\step{a_n}P_n=P'$.

Given a \TCP{} process expression $P$, the transition system $\T(P)=(\Sta_P,\step{}_P,\uparrow_P,\downarrow_P)$ associated with $P$ is defined as follows:
\begin{enumerate}
\item the set of states $\Sta_P$ consists of all process expressions reachable from $P$;
\item the set of transitions $\step{}_P$ is the restriction to $\Sta_P$ of the transition relation defined on all process expressions by the structural operational semantics, i.e., $\Sta_P$ of the $\step{}_P=\step{}\cap(\Sta_P\times\Atau\times\Sta_P)$;
\item ${\uparrow_P}=P$; and
\item the set of final states $\downarrow_P$ consists of all process expressions $Q\in\Sta_P$ such that $Q\downarrow$, i.e., ${\downarrow_P}={\downarrow\cap\Sta_P}$.
\end{enumerate}

We also use the process calculus \SA{} in later sections. It is obtained by excluding the parallel composition operator from \TCP.

The notion of behavioural equivalence has been used extensively in the theory of process calculi.
We first introduce the notion of strong bisimulation~\cite{M1989,P1981}, which does not distinguish $\tau$-transitions from other labelled transitions.

\begin{definition}~\label{def:bisim}
A binary symmetric relation $\R$ on a transition system $(\Sta,\step{},\uparrow,\downarrow)$ is a \emph{strong bisimulation} if, for all states $s,t\in\Sta$, $s\R t$ implies
\begin{enumerate}
 \item if $s\step{a}s'$, then there exist $t'\in\Sta_2$, such that $t\step{a}t'$, and $s'\R t'$;
    \item if $s\downarrow$, then $t\downarrow$.
\end{enumerate}
The states $s$ and $t$ are \emph{strongly bisimilar} (notation: $s\bisim t$) if there exists a strong bisimulation $\R$ s.t. $s\R t$.
\end{definition}

The notion of strong bisimilarity does not take into account the intuition associated with $\tau$ that it stands for unobservable internal activity. To this end, we proceed to introduce the notion of (divergence-preserving) branching bisimilarity, which does treat $\tau$-transitions as unobservable. Divergence-preserving branching bisimilarity in this paper which is the finest behavioural equivalence in van Glabbeek's linear time - branching time spectrum~\cite{Glabbeek1993}. Let $\step{}$ be an $\Atau$-labelled transition relation on a set $\Sta$, and let $a\in\Atau$; we write $s\step{(a)}t$ for the formular ``$s\step{a}t\vee (a=\tau\wedge s=t)$''. Furthermore, we denote the transitive closure of $\step{\tau}$ by $\step{}^{+}$ and the reflexive-transitive closure of $\step{\tau}$ by $\step{}^{*}$.

\begin{definition}
~\label{def:bbisim}
Let $T=(\Sta,\step{},\uparrow,\downarrow)$ be a transition system. A branching bisimulation is a symmetric relation $\R\subseteq\Sta\times\Sta$ such that for all states $s,t\in\Sta$, $s\R t$ implies
\begin{enumerate}
    \item if $s\step{a}s'$, then there exist $t',t''\in\Sta_2$, such that $t\step{}^{*}t''\step{(a)}t'$, $s\R t''$ and $s'\R t'$;
    \item if $s\downarrow$, then there exists $t'$ such that $t\step{}^{*} t'$ and $t'\downarrow$; and
\end{enumerate}
The states $s$ and $t$ are \emph{branching bisimilar} (notation: $s\bbisim t$) if there exists a branching bisimulation $\R$, s.t. $s\R t$.

A branching bisimulation $\R$ is \emph{divergence-preserving} if, for all states $s$ and $t$, $s\R t$ implies
\begin{enumerate}
\setcounter{enumi}{2}
    \item if there exists an infinite sequence $(s_{i})_{i\in\mathbb{N}}$ such that $s=s_{0},\,s_{i}\step{\tau}s_{i+1}$ and $s_{i}\R t$ for all $i\in\mathbb{N}$, then there exists a state $t'$ such that $t\step{}^{+}t'$ and $s_{i}\R t'$ for some $i\in\mathbb{N}$.
    \end{enumerate}
The states $s$ and $t$ are \emph{divergence-preserving branching bisimilar} (notation: $s\bbisimd t$) if there exists a divergence-preserving branching bisimulation $\R$ such that $s\R t$.
\end{definition}

We call the largest divergence-preserving branching bisimulation relation \emph{divergence-preserving branching bisimilarity}. Note that divergence-preserving branching bisimulation relations are equivalence relations~\cite{vGLT2009}.

\begin{definition}
An equivalence relation $\R$ on a process calculus $C$ is called a congruence if $s_i \R t_i$ for $i=1, ..., ar( f )$ implies $f (s_1 ,\ldots, s_{ar( f )})\R f(t_1 ,\ldots, t_{ar(f)})$, where $f$ is an operator of $C$, $ar(f)$ is the arity of $f$, and $s_i,t_i$ are processes defined in $C$.
\end{definition}

Divergence-preserving branching bisimulation relation is not a congruence with respect to most process calculi. A rootedness condition needs to be introduced.

\begin{definition}~\label{def:root}
A divergence-preserving branching bismulation relation $\R$ on a transition system $(\Sta,\step{},\uparrow,\downarrow)$ satisfies \emph{rootedness} condition on a pair of states $s_1,s_2\in\Sta$, if $s_1\R s_2$ and
\begin{enumerate}
\item if $s_1\step{a}s_1'$, then $s_2\step{a}s_2'$ for some $s_2'$ such that $s_1'\R s_2'$;
\item if $s_2\downarrow$, then $s_1\downarrow$.
\end{enumerate}
$s_1$ and  $s_2$ are \emph{rooted divergence-preserving branching bisimilar} (notation: $s_1\rbbisimd s_2$) if there exists a divergence-preserving branching bisimulation $\R$, such that $s_1\R s_2$, and it satisfies rootedness condition on $s_1$ and $s_2$.
\end{definition}

We can extend the above relations ($\bisim,\,\bbisim,\,\bbisimd,$ and $\rbbisimd$) to relations over two transition systems by taking the union of two disjoint transition systems, and two transition systems are bisimilar if their initial states are bisimilar in the union. Namely, for two transition systems $T_1=(\Sta_1,\step{}_1,\uparrow_1,\downarrow_1)$ and $T_2=(\Sta_2,\step{}_2,\uparrow_2,\downarrow_2)$, we make the following pairing on their states. We pair every state $s\in\Sta_1$ with $1$ and every state $s\in\Sta_2$ with $2$. We have $T_i'=(\Sta_i',\step{}_i',\uparrow'_i,\downarrow'_i)$ for $i=1,2$ where $\Sta_i'=\{(s,i)\mid s\in\Sta_i\}$, $\step{}_i'=\{((s,i),a,(t,i))\mid (s,a,t)\in\step{}_i\}$, $\uparrow_i'=(\uparrow_i,i)$, and $\downarrow_i'=\{(s,i)\mid s\in\downarrow_i\}$. We say $T_1\equiv T_2$ if their exists a behaviour equivalence $\equiv$ on $T=(\Sta_1'\cup\Sta_2',{\step{}_1'}\cup{\step{}_2'},\uparrow_1',{\downarrow_1'}\cup{\downarrow_2'})$ such that $\uparrow_1'\equiv\uparrow_2'$.

\section{Transparency}\label{sec:transparency}
Process expressions that have the option to terminate are \emph{transparent} in a sequential context: if $P$ has the
option to terminate and $Q\step{a}Q'$, then $P\cdot Q\step{a}Q'$ even if $P$ can still do transitions. In this section we shall
explain how transparency gives rise to two phenomena that are undesirable in certain circumstances. First, it
facilitates the specification of unboundedly branching behaviour with a guarded recursive specification over
\SA{}.  Second, it gives rise to forgetful stacking of
variables, and as a consequence it is not clear how to specify an always terminating half-counter.

We first discuss process expressions with unbounded branching.
I has been a well-known result from formal language theory that the context-free languages are
exactly the languages accepted by pushdown automata. The process-theoretic formulation of this result is
that every transition system specified by a \SA{} specification is language equivalent to the transition
system associated with a pushdown automaton and, vice versa, every transition system associated with a
pushdown automaton is language equivalent to the transition system associated with some \SA{}
specification. The correspondence fails, however, when language equivalence is replaced by (strong)
bisimilarity. The current best known result is that for every context-free process, there is a pushdown process to simulate it modulo contra simulation~\cite{BCvT2008}. However, we have not succeeded in improving the result to branching bisimulation. The reason is that the context-free processes might have unbounded branching degree when there are unboundedly many transparency process expressions connected by sequential composition. Consider the following process:
\begin{equation*}
X=a.X\cdot Y+b.\one\quad Y=c.\one+\one
\enskip.
\end{equation*}
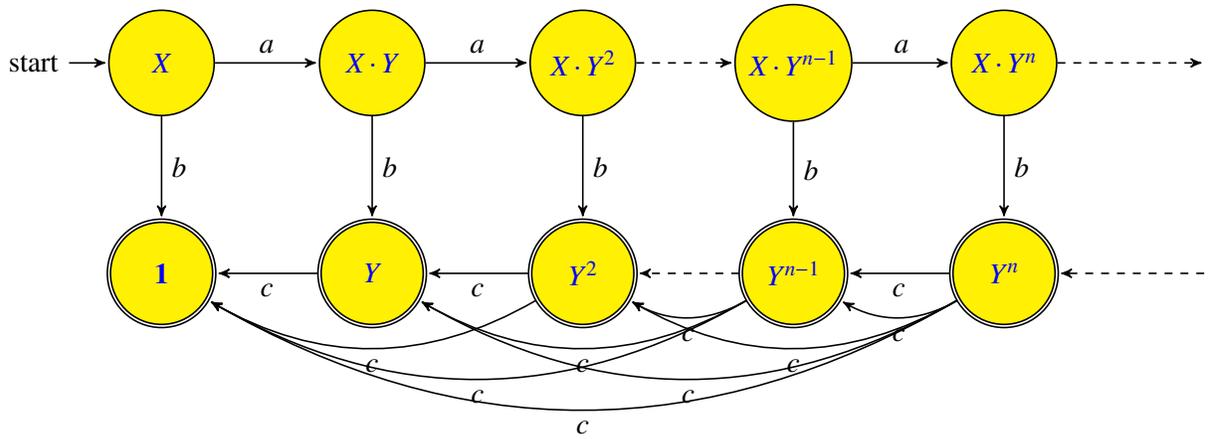
\begin{figure}
\centering
\begin{tikzpicture}[->,>=stealth',shorten >=1pt,auto,node distance=2.8cm,semithick]
  \tikzstyle{every state}=[fill=yellow,draw,text=blue, minimum width = 1.4cm]
    \node[state,initial] (A)                    {$X$};
  \node[state]         (B) [right of=A] {$X\cdot Y$};
  \node[state]         (C) [right of=B] {$X\cdot Y^2$};
  \node[state]         (D) [right of=C] {$X\cdot Y^{n-1}$};
  \node[state]         (E) [right of=D] {$X\cdot Y^{n}$};
  \node[state,accepting]         (F) [below of=E]       {$Y^{n}$};
  \node[state,accepting]         (G) [left of=F]       {$Y^{n-1}$};
\node[state,accepting]         (H) [left of=G]       {$Y^2$};
\node[state,accepting]         (I) [left of=H]       {$Y$};
\node[state,accepting]         (J) [left of=I]       {$\one$};
\node[] (K)[right of=E]{};
\node[] (L)[right of=F]{};
  \path (A) edge              node {$a$} (B)
            edge              node {$b$} (J)
        (B) edge            node {$a$} (C)
            edge              node {$b$} (I)
        (C) edge[dashed]              node {} (D)
            edge              node {$b$} (H)
        (D) edge              node {$a$} (E)
            edge              node {$b$} (G)
        (E) edge              node {$b$} (F)
            edge[dashed]      node {} (K)
        (F) edge              node {$c$} (G)
            edge[bend left]    node {$c$} (H)
            edge[bend left]    node {$c$} (G)
            edge[bend left]    node {$c$} (I)
            edge[bend left]    node {$c$} (J)
        (G) edge[dashed]                node{} (H)
            edge[bend left]    node {$c$} (H)
            edge[bend left]    node {$c$} (I)
            edge[bend left]    node {$c$} (J)
        (H) edge      node {$c$} (I)
            edge[bend left]    node {$c$} (J)
        (I) edge      node {$c$} (J)
        (L) edge[dashed]      node {} (F);
\end{tikzpicture}

\caption{A transition system with unboundedly branching behaviour}~\label{fig:unbounded}
\end{figure}
The transition system of the above process is illustrated in Figure~\ref{fig:unbounded}. Note that $X$ is the initial state, and every state in the second row is a terminating state. For the state $Y^n$, it has $n$ $c$-labelled transitions to $\one,Y,Y^2,\ldots,Y^{n-1}$, respectively. Therefore, every state in this transition system has finitely many transitions leading to distinct states, whereas there is no upper bound on the number of transitions from each state. In this case, we say that this transition system has unboundedly branching degree.

we can prove that the process defined by the \SA{} specification above is not strongly bisimilar to a pushdown process since it is unbounded branching, whereas a pushdown process is always boundedly branching. This correspondence does hold for contra simulation~\cite{BCvT2008}, and that it is an
open problem as to whether the correspondence holds modulo branching bisimilarity. In Section~\ref{sec:cfg}, we show that with the revised composition operator, we can remove such unbounded branching and establish a correspondence between pushdown process and context-free process modulo strong bisimilarity.

Now we proceed to discuss the phenomenon of forgetfulness. A process calculus with iteration and nesting is introduced by Bergstra, Bethke and Ponse~\cite{Bergstra1994,bergstra2001non} in which a binary nesting operator $\nest{}{}$ and a Kleene star operator ${}^{*}$ are added. In this paper, we add these two operators to \TCP. They are intuitively given by the following equations where we use equality to represent strong bisimilarity:
\begin{equation*}
P^{*}= P\cdot P^{*}+\one\quad
\nest{P_1}{P_2}= P_1\cdot (\nest{P_1}{P_1})\cdot P_1+P_2
\enskip
\end{equation*}
We give the operational semantics of these two operators in Figure~\ref{fig:semantics-iteration-nesting}.
\begin{figure}
\fbox{
\begin{minipage}[t]{1\textwidth}
\begin{eqnarray*}
&\inference{}{P^{*}\downarrow}\quad
\inference{P\step{a}P'}{P^{*}\step{a}P'\cdot P^{*}}\\
&\inference{P_1\step{a}P_1'}{\nest{P_1}{P_2}\step{a}P_1'\cdot\nest{P_1}{P_2}\cdot P_1}\quad
\inference{P_2\step{a}P_2'}{\nest{P_1}{P_2}\step{a}P_2'}\quad
\inference{P_2\downarrow}{\nest{P_1}{P_2}\downarrow}
\enskip.
\end{eqnarray*}
\end{minipage}
}
\caption{The operational semantics of nesting and iteration}~\label{fig:semantics-iteration-nesting}
\end{figure}

Bergstra et al. show how one can specify a half counter using iteration and nesting, which then allows them to
conclude that the behaviour of a Turing machine can be simulated in the calculus with iteration and nesting~\cite{Bergstra1994,bergstra2001non}.

The half counter is specified as follows:
\begin{eqnarray*}
CC_n &=& \mathit{a}.CC_{n+1}+\mathit{b}.BB_{n}\,(n\in\mathbb{N})\\
BB_n &=& \mathit{a}.BB_{n-1}\,(n\geq 1)\\
BB_0 &=& \mathit{c}.CC_0
\enskip.
\end{eqnarray*}
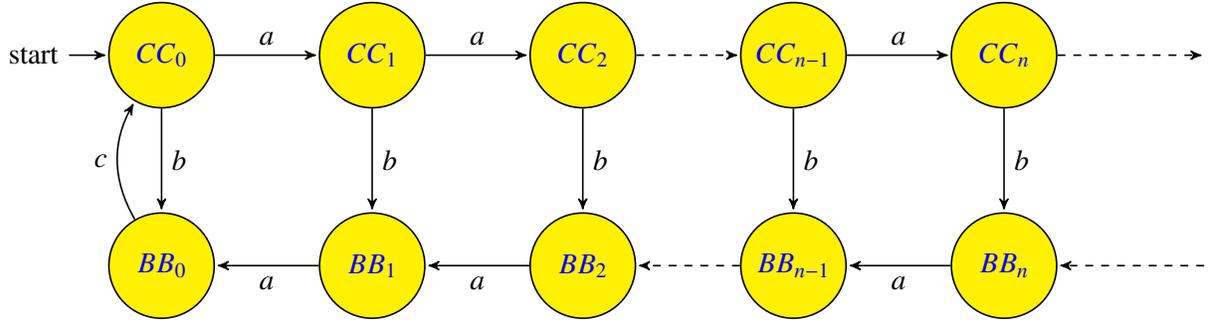
\begin{figure}
\centering
\begin{tikzpicture}[->,>=stealth',shorten >=1pt,auto,node distance=2.8cm,semithick]
  \tikzstyle{every state}=[fill=yellow,draw,text=blue, minimum width = 1.4cm]
    \node[state,initial] (A)                    {$CC_0$};
  \node[state]         (B) [right of=A] {$CC_1$};
  \node[state]         (C) [right of=B] {$CC_2$};
  \node[state]         (D) [right of=C] {$CC_{n-1}$};
  \node[state]         (E) [right of=D] {$CC_{n}$};
  \node[state]         (F) [below of=E]       {$BB_{n}$};
  \node[state]         (G) [left of=F]       {$BB_{n-1}$};
\node[state]         (H) [left of=G]       {$BB_2$};
\node[state]         (I) [left of=H]       {$BB_1$};
\node[state]         (J) [left of=I]       {$BB_0$};
\node[] (K)[right of=E]{};
\node[] (L)[right of=F]{};
  \path (A) edge              node {$a$} (B)
            edge              node {$b$} (J)
        (B) edge            node {$a$} (C)
            edge              node {$b$} (I)
        (C) edge[dashed]              node {} (D)
            edge              node {$b$} (H)
        (D) edge              node {$a$} (E)
            edge              node {$b$} (G)
        (E) edge              node {$b$} (F)
            edge[dashed]      node {} (K)
        (F) edge              node {$a$} (G)
        (G) edge[dashed]                node{} (H)
        (H) edge      node {$a$} (I)
        (I) edge      node {$a$} (J)
        (L) edge[dashed]      node {} (F)
        (J) edge [bend left]            node{$c$} (A);
\end{tikzpicture}
\caption{The transition system of a half counter}~\label{fig:halfcounter}
\end{figure}
The behaviour of a half counter is illustrated in Figure~\ref{fig:halfcounter}. The initial state is $CC_0$. From $CC_0$, it is able to make an arbitrary number of $a$ transitions. At some point, it stops counting with a $b$-labelled transition, and then makes the same number of $a$-labelled transitions to the state $BB_0$. In state $BB_0$, a zero testing transition is labelled by $c$ which leads back to the state $CC_0$.

An implementation in \TCP{} with iteration and nesting is provided in~\cite{bergstra2001non} as follows:
\begin{equation*}
HCC=((\nest{a}{b})\cdot c)^{*}
\enskip.
\end{equation*}

It is straightforward to establish that $((\nest{a}{b})\cdot a^{n}\cdot c) \cdot HCC$ is equivalent to $CC_n$ for all $n\geq 1$ modulo strong bisimilarity, and $(a^{n}\cdot c)\cdot HCC$ is equivalent to $BB_n$ for all $n\in\mathbb{N}$ modulo strong bisimilarity.

In a context with intermediate termination, one may wonder if it is possible to generalize their result. It is,
however, not clear how to specify an always terminating half counter. At least, a naive generalisation of the
specification of Bergstra et al. does not do the job. The culprit is forgetfulness.
We define a half counter that terminates in every state as follows:
\begin{eqnarray*}
C_n &=& \mathit{a}.C_{n+1}+\mathit{b}.B_{n}+\one\,(n\in\mathbb{N})\\
B_n &=& \mathit{a}.B_{n-1}+\one\,(n\geq 1)\\
B_0 &=& \mathit{c}.C_0+\one
\enskip.
\end{eqnarray*}
The following implementation is no longer valid:
\begin{equation*}
HC=(\nest{\mathit{(a+\one)}}{\mathit{(b+\one)}}\cdot (c+\one))^{*}
\enskip.
\end{equation*}
Note that due to transparency, $((a+\one)^{n}\cdot (c+\one))\cdot HC$ is no longer equivalent to $B_n$ modulo any behavioural equivalence for $n>1$ since $B_n$ only has an $a$-labelled transition to $B_{n-1}$ whereas the other process has at least $n+1$ transitions leading to $HC,\,(c+\one)\cdot HC,\,(a+\one)\cdot (c+\one)\cdot HC,\,\ldots,\,(a+\one)^{n-1}\cdot (c+\one)\cdot HC$, respectively. This process may choose to ``forget'' the transparent process expressions that has been stacked with sequential composition operator. The forgetfulness leads to the failure to implement a terminating half counter.

In Section~\ref{sec:Termination}, we show that with the revised semantics, the forgetfulness would be eliminated. Therefore, we show that we are able to implement a terminating half counter with the revised semantics and we shall provide a reactively Turing powerful process calculus. 
\section{A Revised Semantics of the Sequential Composition Operator}\label{sec:sequential}
We propose a calculus \TCPS{} with a new sequential composition operator. Its syntax obtained by replacing the sequential composition operator $\cdot$ by $\bullet$ in the syntax of \TCP{}.
Note that we also use the abbreviation of $P^{n}$ as we did for the standard version of the sequential composition operator.

The structural operational semantics of $\bullet$ is defined in Figure~\ref{fig:revised-semantics-sequential}.
\begin{figure}
\fbox{
\begin{minipage}[t]{1\textwidth}
\begin{eqnarray*}
& \inference{P_1\downarrow\quad P_2\downarrow}{P_1\bullet P_2\downarrow}\quad
\inference{P_1\step{a}P_1'}{P_1\bullet P_2\step{a}P_1'\bullet P_2}\quad \inference{P_1\downarrow\quad P_2\step{a}P_2'\quad P_1\not{\step{}}}{P_1\bullet P_2\step{a}P_2'}
\enskip.
\end{eqnarray*}
\end{minipage}
}
\caption{The revised semantics of sequential composition}~\label{fig:revised-semantics-sequential}
\end{figure}
 We use $P\not{\step{}}$ to denote that there does not exist a closed term $P'$ such that $P\step{a}P'$ is derivable from the operational rules.
With the revised semantics, processes with intermediate termination (option to terminate and option to do an
action) lose their transparency in a sequential context. As a consequence, the branching degree of a
context-free process is bounded and sequential compositions may have the option to terminate, without
being forgetful.

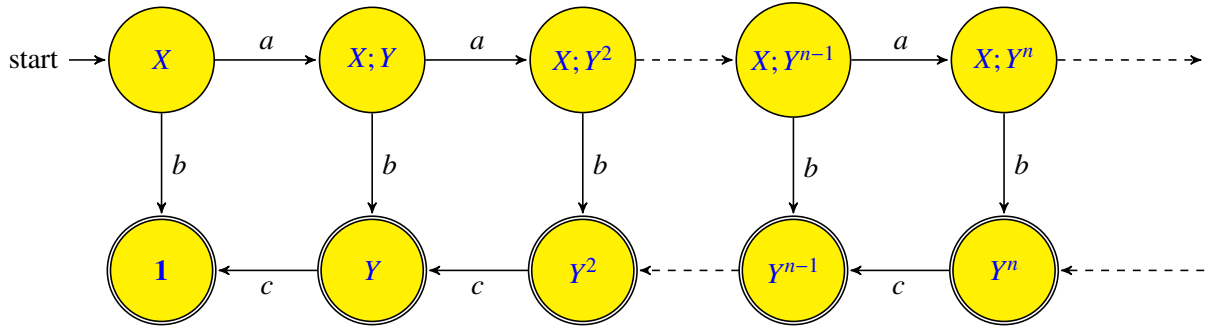
\begin{figure}
\centering
\begin{tikzpicture}[->,>=stealth',shorten >=1pt,auto,node distance=2.8cm,semithick]
  \tikzstyle{every state}=[fill=yellow,draw,text=blue, minimum width = 1.4cm]
    \node[state,initial] (A)                    {$X$};
  \node[state]         (B) [right of=A] {$X\bullet Y$};
  \node[state]         (C) [right of=B] {$X\bullet Y^2$};
  \node[state]         (D) [right of=C] {$X\bullet Y^{n-1}$};
  \node[state]         (E) [right of=D] {$X\bullet Y^{n}$};
  \node[state,accepting]         (F) [below of=E]       {$Y^{n}$};
  \node[state,accepting]         (G) [left of=F]       {$Y^{n-1}$};
\node[state,accepting]         (H) [left of=G]       {$Y^2$};
\node[state,accepting]         (I) [left of=H]       {$Y$};
\node[state,accepting]         (J) [left of=I]       {$\one$};
\node[] (K)[right of=E]{};
\node[] (L)[right of=F]{};
  \path (A) edge              node {$a$} (B)
            edge              node {$b$} (J)
        (B) edge            node {$a$} (C)
            edge              node {$b$} (I)
        (C) edge[dashed]              node {} (D)
            edge              node {$b$} (H)
        (D) edge              node {$a$} (E)
            edge              node {$b$} (G)
        (E) edge              node {$b$} (F)
            edge[dashed]      node {} (K)
        (F) edge              node {$c$} (G)
        (G) edge[dashed]                node{} (H)
        (H) edge      node {$c$} (I)
        (I) edge      node {$c$} (J)
        (L) edge[dashed]      node {} (F);
\end{tikzpicture}
\caption{The transition system in the revised semantics}\label{fig:bounded}
\end{figure}

Let us revisit the example in Section~\ref{sec:transparency}. We rewrite it with the revised sequential composition operator:
\begin{eqnarray*}
X&=&a.X\bullet Y+b.\one\\
Y&=&c.\one+\one
\enskip.
\end{eqnarray*}
Its transition system is illustrated in Figure~\ref{fig:bounded}. Every state in the transition system now has a bounded branching degree. For instance, a transition from $Y^{5}$ to $Y^{2}$ is abandoned because $Y$ has a transition and only the transition from the first $Y$ in the sequential composition is allowed.

Congruence is an important property to fit a behavioural equivalence into an axiomatic framework. We show that in the revised semantics, $\rbbisimd$ is a congruence. Note that the congruence property can also be inferred from a recent result of Fokknik, van Glabbeek and Luttik~\cite{FvGL2017}.
\begin{theorem}~\label{thm:congruence}
$\rbbisimd$ is a congruence with respect to \TCPS.
\end{theorem}
\arx{
\begin{proof}
We use the following facts:
\begin{enumerate}
\item  Rooted divergence-preserving branching bisimilarity
is a rooted divergence-preserving branching
bisimulation relation; and
\item rooted divergence-preserving branching bisimilarity
is a subset of divergence-preserving branching
bisimilarity.
\end{enumerate}
We show that $\rbbisimd$ is compatible for each operator $a.,+,\bullet,\parallel$.
\begin{enumerate}
\item Suppose that $P\rbbisimd Q$, we show that $a.P\rbbisimd a.Q$. To this end, we verify that $\R=\{(a.P,a.Q)\mid P\rbbisimd Q\}\cup{\rbbisimd}$ is a rooted divergence-preserving branching bisimulation relation.

    To prove that the pair $(a.P, a.Q)$ with $P$ rooted divergence-preserving branching bisimilar to Q satisfies
condition 1 of Definition~\label{def:root}, suppose that $a.P\step{b}P'$. Then, according to the operational semantics, $b=a$
and $P'=P$. By the operational semantics, we also have that $a.Q\step{a} Q$ and, by assumption, $P$ and $Q$ are
divergence-preserving branching bisimilar.

    For the termination condition, it is trivially satisfied since both processes do not terminate. The divergence-preserving condition is also satisfied since only an $a$-labelled transition is allowed from both processes.

\item Suppose that $P_1\rbbisimd Q_1$ and $P_2\rbbisimd Q_2$, we show that $P_1+P_2\rbbisimd Q_1+Q_2$. To this end, we verify that $\R=\{(P_1+P_2,Q_1+Q_2)\mid P_1\rbbisimd Q_1,\,P_2\rbbisimd Q_2\}\cup{\rbbisimd}$ is a rooted divergence-preserving branching bisimulation relation.

    Suppose that $P_1+P_2\step{a}P'$; then we have $P_1\step{a}P'$ or $P_2\step{a}P'$. We only consider the first case.
    Since $P_1\rbbisimd Q_1$, we have $Q_1\step{a}Q'$ with $P'\bbisimd Q'$. Then we have $Q_1+Q_2\step{a}Q'$ with $P'\bbisimd Q'$. The same argument holds for the symmetrical case.

    If $P_1+P_2\downarrow$ , then we have either $P_1\downarrow$ or $P_2\downarrow$. Without loss of generality, we suppose that $P_1\downarrow$. Since $P_1\rbbisimd Q_1$, we have $Q_1\downarrow$. Therefore, $Q_1+Q_2\downarrow$.

    Moreover, we verify that the divergence preservation condition is satisfied.

    Hence, $\R$ is a rooted divergence-preserving branching bisimulation relation.

\item Suppose that $P_1\rbbisimd Q_1$ and $P_2\rbbisimd Q_2$, we show that $[P_1\parallel P_2]_{\C'}\rbbisimd [Q_1\parallel Q_2]_{\C'}$. To this end, we verify that $\R=\{([P_1\parallel P_2]_{\C'},[Q_1\parallel Q_2]_{\C'})\mid P_1\rbbisimd Q_1,\,P_2\rbbisimd Q_2\}\cup{\rbbisimd}$ is a rooted divergence-preserving branching bisimulation relation.

    We first show that $\R'=\{([P_1\parallel P_2]_{\C'},[Q_1\parallel Q_2]_{\C'})\mid P_1\bbisimd Q_1,\,P_2\bbisimd Q_2\}\cup{\bbisimd}$ is a divergence-preserving branching bisimulation.

    Suppose that $[P_1\parallel P_2]_{\C'}\step{a}P'$; then we distinguish several cases.
    \begin{enumerate}
    \item If $P_1\step{a}P_1'\,a\notin \I_{\C'}$ and $P'=[P_1'\parallel P_2]_{\C'}$, then, since $P_1\bbisimd Q_1$, we have $Q_1\step{}^{*}Q_1''\step{a}Q_1'$ with $P_1'\bbisimd Q_1'$ and $P_1\bbisimd Q_1''$. Then we have $[Q_1\parallel Q_2]_{\C'}\step{}^{*}[Q_1''\parallel Q_2]_{\C'}\step{a}[Q_1'\parallel Q_2]_{\C'}$ with $P_1\bbisimd Q_1''$, $P_1'\bbisimd Q_1'$ and $P_2\bbisimd Q_2$. Thus we have $([P_1'\parallel P_2]_{\C'},[Q_1'\parallel Q_2]_{\C'})\in\R'$ and $([P_1\parallel P_2]_{\C'},[Q_1''\parallel Q_2]_{\C'})\in\R'$.

    \item If $P_1\step{c?d}P_1',\,P_2\step{c!d}P_2'$ and $ c\in\C'$, then $[P_1\parallel P_2]_{\C'}\step{\tau}[P_1'\parallel P_2']_{\C'}$. Since $P_1\bbisimd Q_1$ and $P_2\bbisimd Q_2$, we have $Q_1\step{}^{*}Q_1''\step{c?d}Q_1',\,Q_2\step{}^{*}Q_2''\step{c!d}Q_2'$ with $P_1'\bbisimd Q_1'$ , $P\bbisimd Q_1''$, $P_2'\bbisimd Q_2'$, and $P_2\bbisimd Q_2''$. Then we have $[Q_1\parallel Q_2]_{\C'}\step{}^{*}[Q_1''\parallel Q_2'']_{\C'}\step{\tau} [Q_1'\parallel Q_2']_{\C'}$ with $P_1\bbisimd Q_1''$, $P_2\bbisimd Q_2''$, $P_1'\bbisimd Q_1'$ and $P_2'\bbisimd Q_2'$. Thus we have $([P_1\parallel P_2]_{\C'},[Q_1''\parallel Q_2'']_{\C'})\in\R'$ and $([P_1'\parallel P_2']_{\C'},[Q_1'\parallel Q_2']_{\C'})\in\R'$.
    \end{enumerate}

    If $[P_1\parallel P_2]_{\C'}\downarrow$, then we have $P_1\downarrow$ and $P_2\downarrow$. Since $P_1\bbisimd Q_1$ and $P_2\bbisimd Q_2$, we have $Q_1\step{}^{*}Q_1'\downarrow$ and $Q_2\step{}^{*}Q_2'\downarrow$ for some $Q_1'$ and $Q_2'$. Therefore, $[Q_1\parallel Q_2]_{\C'}\step{}^{*}[Q_1'\parallel Q_2']_{\C'}\downarrow$.

    Hence, we have $\R'$ is a divergence-preserving branching bisimulation relation.

    Now we show that $\R$ is a rooted divergence-preserving branching bisimulation.
    Suppose that $[P_1\parallel P_2]_{\C'}\step{a}P'$; then we distinguish several cases.
    \begin{enumerate}
    \item If $P_1\step{a}P_1'\,a\notin \I_{\C'}$ and $P'=[P_1'\parallel P_2]_{\C'}$, then, since $P_1\rbbisimd Q_1$, we have $Q_1\step{a}Q_1'$ with $P_1'\bbisimd Q_1'$. Then we have $[Q_1\parallel Q_2]_{\C'}\step{a} [Q_1'\parallel Q_2]_{\C'}$ with $P_1'\bbisimd Q_1'$ and $P_2\bbisimd Q_2$. Thus we have $[P_1'\parallel P_2]_{\C'}\bbisimd [Q_1'\parallel Q_2]_{\C'}$.

    \item If $P_1\step{c?d}P_1',\,P_2\step{c!d}P_2'$ and $ c\in\C'$, then $[P_1\parallel P_2]_{\C'}\step{\tau}[P_1'\parallel P_2']_{\C'}$. Since $P_1\rbbisimd Q_1$ and $P_2\rbbisimd Q_2$, we have $Q_1\step{c?d}Q_1',\,Q_2\step{c!d}Q_2'$ with $P_1'\bbisimd Q_1'$ and $P_2'\bbisimd Q_2'$. Then we have $[Q_1\parallel Q_2]_{\C'}\step{\tau} [Q_1'\parallel Q_2']_{\C'}$ with $P_1'\bbisimd Q_1'$ and $P_2'\bbisimd Q_2'$. Thus we have $[P_1'\parallel P_2']_{\C'}\bbisimd [Q_1'\parallel Q_2']_{\C'}$.
    \end{enumerate}

    If $[P_1\parallel P_2]_{\C'}\downarrow$, then we have $P_1\downarrow$ and $P_2\downarrow$. Since $P_1\rbbisimd Q_1$ and $P_2\rbbisimd Q_2$, we have $Q_1\downarrow$ and $Q_2\downarrow$. Therefore, $[Q_1\parallel Q_2]_{\C'}\downarrow$.

    Moreover, we verify that the divergence preservation condition is satisfied.

    Hence, we have $\R$ is a rooted divergence-preserving branching bisimulation relation.
\item  Suppose that $P_1\rbbisimd Q_1$ and $P_2\rbbisimd Q_2$, we show that $P_1\bullet P_2\rbbisimd Q_1\bullet Q_2$. To this end, we verify that $\R=\{(P_1\bullet P_2,Q_1\bullet Q_2)\mid P_1\rbbisimd Q_1,\,P_2\rbbisimd Q_2\}\cup{\rbbisimd}$ is a rooted divergence-preserving branching bisimulation relation.

We first show that $\R'=\{(P_1\bullet P_2,Q_1\bullet Q_2)\mid P_1\bbisimd Q_1,\,P_2\rbbisimd Q_2\}\cup{\bbisimd}$ is a divergence-preserving branching bisimulation relation.

 Suppose that $P_1\bullet P_2\step{a}P'$; then we distinguish several cases.
    \begin{enumerate}
    \item If $P_1\step{a} P_1'$, then $P'=P_1'\bullet P_2$. Since $P_1\bbisimd Q_1$, we have $Q_1\step{}^{*}Q_1''\step{a}Q_1'$ with $P_1'\bbisimd Q_1'$ and $P_1\bbisimd Q_1''$. Then we have $Q_1\bullet Q_2 \step{}^{*}Q_1''\bullet Q_2\step{a}Q_1'\bullet Q_2$ with $P_1\bbisimd Q_1''$, $P_1'\bbisimd Q_1'$, and $P_2\rbbisimd Q_2$. Thus, we have $(P_1'\bullet P_2, Q_1'\bullet Q_2)\in\R'$ and $(P_1\bullet P_2,Q_1''\bullet Q_2)\in\R'$.
    \item If $P_1\downarrow,\, P_2\step{a}P_2'$ and $P_1\not{\step{}}$. Since $P_1\bbisimd Q_1$ and $P_2\rbbisimd Q_2$, we have $Q_1\step{}^{*}Q_1'\downarrow$, $Q_1'\not{\step{}}$ for some $Q_1'$ with $P_1\bbisimd Q_1'$, and $Q_2\step{a}Q_2'$, with $P_2'\bbisimd Q_2'$. Then, we have $Q_1\bullet Q_2\step{}^{*}Q_1'\bullet Q_2\step{a} Q_2'$ with $P_2'\bbisimd Q_2'$ and $P_1\bbisimd Q_1'$. Thus we have $(P_2',Q_2')\in\R'$ and $(P_1\bullet P_2,Q_1'\bullet Q_2)\in\R$.
    \end{enumerate}

    If $P_1\bullet P_2\downarrow$, then we have $P_1\downarrow$ and $P_2\downarrow$. Since $P_1\bbisimd Q_1$ and $P_2\rbbisimd Q_2$, we have $Q_1\step{}^{*}Q_1'\downarrow$ for some $Q_1'$ and $Q_2\downarrow$. Therefore, $Q_1\bullet Q_2\step{}^{*}Q_1'\bullet Q_2\downarrow$.

    Moreover, we verify that the divergence preservation condition is satisfied.

    Hence, we have $\R$ is a divergence-preserving branching bisimulation relation.

Now we show that $\R$ is a rooted divergence-preserving branching bisimulation relation.

    We suppose that $P_1\bullet P_2\step{a}P'$, we distinguish several cases:
    \begin{enumerate}
    \item If $P_1\step{a} P_1'$, then $P'=P_1'\bullet P_2$. Since $P_1\rbbisimd Q_1$, we have $Q_1\step{a}Q_1'$ with $P_1'\bbisimd Q_1'$. Then we have $Q_1\bullet Q_2 \step{a}Q_1'\bullet Q_2$ with $P_1'\bbisimd Q_1'$ and $P_2\rbbisimd Q_2$. Thus, we have $P_1'\bullet P_2\bbisimd Q_1'\bullet Q_2$.
    \item If $P_1\downarrow,\, P_2\step{a}P_2'$ and $P_1\not{\step{}}$. Since $P_1\rbbisimd Q_1$ and $P_2\rbbisimd Q_2$, we have $Q_1\downarrow$, $Q_2\step{a}Q_2'$, with $P_2'\bbisimd Q_2'$, and $Q_1\not{\step{}}$. Then, we have $Q_1\bullet Q_2\step{a} Q_2'$ with $P_2'\bbisimd Q_2'$.
    \end{enumerate}

    If $P_1\bullet P_2\downarrow$, then we have $P_1\downarrow$ and $P_2\downarrow$. Since $P_1\rbbisimd Q_1$ and $P_2\rbbisimd Q_2$, we have $Q_1\downarrow$ and $Q_2\downarrow$. Therefore, $Q_1\bullet Q_2\downarrow$.

    Moreover, we verify that the divergence preservation condition is satisfied.

    Hence, we have $\R$ is a rooted divergence-preserving branching bisimulation relation.
\end{enumerate}
\end{proof}
}

We also define a version of \TCP{} with iteration and nesting (\TCPN{}) in the revised semantics.
By removing recursive specification and adding a non-regular iterator, we get \TCPN{} as a variation of \TCPS{} with two additional operators: $P^{*},\nest{P_1}{P_2}$.
\delete{
We extend the calculus with the following operators:
\begin{equation*}
P^{*},\pushd{P_1}{P_2},\backf{P_1}{P_2},\nest{P_1}{P_2}
\end{equation*}

These operations are defined by
\begin{eqnarray*}
\pushd{x}{y}&=& x(\pushd{x}{y})(\pushd{x}{y})+y\\
\backf{x}{y}&=& x(\backf{x}{y})y+y\\
\nest{x}{y}&=& x(\nest{x}{y})x+y
\end{eqnarray*}
}
The operational semantics is defined in Figure~\ref{fig:revised-semantics-iteration-nesting}.
\begin{figure}
\fbox{
\begin{minipage}[t]{1\textwidth}
\begin{eqnarray*}
&\inference{}{P^{*}\downarrow}\quad
\inference{P\step{a}P'}{P^{*}\step{a}P'\bullet P^{*}}\\
\delete{&\inference{P_1\step{a}P_1'}{\pushd{P_1}{P_2}\step{a}P_1'\bullet\pushd{P_1}{P_2}\bullet\pushd{P_1}{P_2}}\quad
\inference{P_2\step{a}P_2'}{\pushd{P_1}{P_2}\step{a}P_2'}\quad
\inference{P_2\downarrow}{\pushd{P_1}{P_2}\downarrow}\\
&\inference{P_1\step{a}P_1'}{\backf{P_1}{P_2}\step{a}P_1'\bullet\backf{P_1}{P_2}\bullet P_2}\quad
\inference{P_2\step{a}P_2'}{\backf{P_1}{P_2}\step{a}P_2'}\quad
\inference{P_2\downarrow}{\backf{P_1}{P_2}\downarrow}\\}
&\inference{P_1\step{a}P_1'}{\nest{P_1}{P_2}\step{a}P_1'\bullet\nest{P_1}{P_2}\bullet P_1}\quad
\inference{P_2\step{a}P_2'}{\nest{P_1}{P_2}\step{a}P_2'}\quad
\inference{P_2\downarrow}{\nest{P_1}{P_2}\downarrow}
\end{eqnarray*}
\end{minipage}
}
\caption{The revised semantics of iteration and nesting}~\label{fig:revised-semantics-iteration-nesting}
\end{figure} 
\section{Context-free Processes and Pushdown Processes}\label{sec:cfg}
The relationship between context-free processes and pushdown processes has been studied extensively in the literature~\cite{BCvT2008}.

We consider the process calculus \emph{Theory of Sequential Processes} (\SAS) with the revised semantics of sequential composition operator as follows:
\begin{equation*}
P=\nil\mid\one\mid a.P\mid P_1 \bullet P_2\mid P_1+P_2\mid N
\enskip.
\end{equation*}

We give the definition of context-free processes as follows:

\begin{definition}~\label{def:cfp}
A \emph{context-free process} is the bisimulation equivalence class of the transition system generated by a finite guarded recursive specification over Sequential Algebra \SAS.
\end{definition}

Note that every context-free process can be rewritten into a Greibach normal form. In this paper, we only consider all the context-free processes (guarded recursive specifications) written in Greibach normal form:
\begin{equation*}
X=\sum_{i\in I_X}\alpha_i.\xi_i(+\one)
\enskip.
\end{equation*}

In this form, every right-hand side of every equation consists of a number of summands, indexed by a finite set $I_X$ (the empty sum is $\nil$), each of which is $\one$, or of the form $\alpha_i.\xi_i$, where $\xi_i$ is the sequential composition of a number of names (the empty sequence is $\one$). \delete{We define $I$ as the multiset resulting of the union of all index sets. For a recursive specification in Greibach normal form, every state of the transition system is given by a sequence of names. Note that we can take the index sets associated with the names to be disjoint, so that we can define a function $V : I\rightarrow \V$ that gives, for any index that occurs somewhere in the specification, the name of the equation in which it occurs.}

We shall show that every context-free process is equivalent to a pushdown automata modulo strong bisimilarity. The notion of pushdown automata is defined as follows:

\begin{definition}~\label{def:pda}
A pushdown automaton (\PDA) is a $7$-tuple $(\Sta,\Sigma,\D,\step{},\uparrow,Z,\downarrow)$, where
\begin{enumerate}
\item $\Sta$ is a finite set of \emph{states},
\item $\Sigma$ is a finite set of \emph{input symbols},
\item $\D$ is a finite set of \emph{stack symbols},
\item $\step{}\subseteq \Sta\times\Sigma\times\D\times\Sigma^{*}\Sta$ is a finite \emph{transition relation}, (we write $s\step{a[d/\delta]}t$ for $(s,d,a,\delta,t)\in\step{}$),
\item $\uparrow\in\Sta$ is the \emph{initial state},
\item $Z\in\D$ is the \emph{initial stack symbol}, and
\item $\downarrow\subseteq\Sta$ is a set of \emph{accepting states}.
\end{enumerate}
\end{definition}

We use a sequence of stack symbols $\delta\in\D^{*}$ to represent the contents of a stack. We associate with every pushdown automaton a labelled transition system. We call the transition system associated with a pushdown automaton a pushdown process.
\begin{definition}~\label{def:lts-pda}
   Let $\M=(\Sta,\Sigma,\D,\step{},\uparrow,Z,\downarrow)$ be a \PDA. The \emph{pushdown process} $\T(\M)=(\Sta_{\T},\step{}_{\T},\uparrow_{\T},\downarrow_{\T})$ associated $\M$ is defined as follows:
\begin{enumerate}
\item its set of states is the set $\Sta_{\T}= \{(s,\delta)\mid s\in\Sta,\delta\in\D^{*}\}$ of all configurations of $\M$;
\item its transition relation $\step{}_{\T}\subset \Sta_{\T}\times \Atau\times \Sta_{\T}$ is the least relation satisfying, for all $a\in\Sigma$, $d\in\D$, $\delta,\delta'\in\D^{*}$: $(s,d\delta)\step{a}_{\T}(t,\delta'\delta)$ iff $s\step{a[d/\delta']}t$.
\item its initial state is the configuration $\uparrow_{\T}=(\uparrow,Z)$, and
\item its set of final states is the set $\downarrow_{\T}=\{(s,\delta)\mid s\in\Sta,\,s\downarrow,\,\delta\in\D^{*}\}$.
\end{enumerate}
\end{definition}

Consider a context-free process in Greibach normal form defined by a set of names $\V=\{X_0,X_1,\ldots,X_m\}$ where
\begin{equation*}
X_j=\sum_{i\in I_{X_j}}\alpha_{ij}.\xi_{ij}(+\one)
\enskip,
\end{equation*}
and $X_0$ is the initial state.

We introduce the following functions for a sequence of names $\xi$：
\begin{enumerate}
\item $\mathit{length}:\V^{*}\rightarrow\mathbb{N}$, $\length{\xi}$ computes the length of $\xi$;
\item $\mathit{get}:\V^{*}\times\mathbb{N}\rightarrow \V$, $\get{\xi}{i}$ computes the $i$-th name of $\xi$;
\item $\mathit{suffset}:\V^{*}\times\mathbb{N}\rightarrow 2^{\lvert \V\rvert}$, $\suffset{\xi}{i}=\{\get{\xi}{j}\mid j=i+1,\ldots\length{\xi}\}$ computes the set that contains all the names in the suffix which starts from the $i$-th name of $\xi$.
\end{enumerate}
Next we define a \PDA{} $\M=(\Sta,\Sigma,\D,\step{},\uparrow,Z,\downarrow)$ to simulate the transition system associated with $X_0$ as follows:
\begin{enumerate}
\item $\Sta=\{s_D\mid D\subseteq \V\}$;
\item $\Sigma=\Atau$;
\item $\D=\V\cup\{X^{\dagger}\mid X\in\V\}$;
\item the set of transitions $\step{}$ is defined as follows:
\begin{eqnarray*}
\step{}&=&\{(s_D,X^{\dagger}_{j},\alpha_{ij},\delta(s_D,X^{\dagger}_{j},\xi_{ij}),s_{D(s_D,X^{\dagger}_{j},\xi_{ij})})\mid i\in I_{X_j},\,j=1,\ldots,n,\,D\subset\V\}\\
&\cup&\{(s_D,X_j,\alpha_{ij},\delta(s_D,X_j,\xi_{ij}),s_{D(s_D,X_j,\xi_{ij})})\mid i\in I_{X_j},\,j=1,\ldots,n,\,D\subset\V\}
\enskip.
\end{eqnarray*}
where $\delta(s_D,X^{\dagger}_{j},\xi_{ij})$ is a string of length $\length{\xi_{ij}}$ defined as follows: for $k=1,\ldots, \length{\xi_{ij}}$, we let $X_k=\get{\xi_{ij}}{k}$,
\begin{enumerate}
 \item if $X_k\notin (D/\{X_j\})\cup\suffset{\xi_{ij}}{k}$, then the $k$-th symbol of $\delta(s_D,X^{\dagger}_{j},\xi_{ij})$ is $X^{\dagger}_{k}$,
\item otherwise, the $k$-th symbol of $\delta(s_D,X^{\dagger}_{j},\xi_{ij})$ is $X_k$,
\end{enumerate}
$\delta(s_D,X_j,\xi_{ij})$ is a string of length $\length{\xi_{ij}}$ defined as follows: for $k=1,\ldots,\length{\xi_{ij}}$, we let $X_k=\get{\xi_{ij}}{k}$,
\begin{enumerate}
 \item if $X_{k}\notin D\cup\suffset{\xi_{ij}}{k}$, then the $k$-th symbol of $\delta(s_D,X_j,\xi_{ij})$ is $X^{\dagger}_k$,
\item otherwise, the $k$-th symbol of $\delta(s_D,X_j,\xi_{ij})$ is $X_k$, and
\end{enumerate}
we also define $D(s_D,X^{\dagger}_j,\xi_{ij})=(D/\{X_j\})\cup\suffset{\xi_{ij}}{0}$ and $D(s_D,X_j,\xi_{ij})=D\cup\suffset{\xi_{ij}}{0}$;
\item $\uparrow=s_{\{X_0\}}$;
\item $Z=X^{\dagger}_{0}$;
\item $\downarrow=\{s_D\mid \mbox{if for all}\,X\in D, X\downarrow\}$.
\end{enumerate}

We observe that every process expression $\xi$ is simulated by a configuration of $\M$ such that the sequence of names in $\xi$ is stored in the stack. The first appearance of every name from the bottom of the stack is marked with $\dagger$. The state is marked by a set that contains all the names in $\xi$. A state is terminating if and only if all the names in the set that marks the state are terminating. We show the following result:
\begin{lemma}~\label{lemma:cfp-pda}
$\T(X_0)\bisim \T(\M)$.
\end{lemma}
\arx{
\begin{proof}
We first define an auxiliary function $\mathit{stack}:\V^{*}\rightarrow \D^{*}$ as follows: given $\xi\in\V^{*}$, for $k=1,\ldots,\length{\xi}$, we let $X_k=\get{\xi}{k}$,
\begin{enumerate}
\item  if $X_k\notin\suffset{\xi}{k}$, then the $k$-th the element of $\stack{\xi}$ is $X^{\dagger}_{k}$,
\item otherwise, the $k$-the the element of $\stack{\xi}$ is $X_{k}$,
\end{enumerate}
Note that $\stack{X\xi}$ and $\stack{\xi}$ share the same suffix of length $\length{\xi}$.

We show that the following relation:
\begin{equation*}
\R=\{(\xi,(s_{\suffset{\xi}{0}},\mathit{stack}(\xi)))\mid \xi\in\V^{*}\}
\enskip,
\end{equation*}
is a strong bisimulation.

We rewrite $\xi$ as $X_j\xi'$, then it has the following transitions:
\begin{equation*}
X_j\xi'\step{\alpha_{ij}} \xi_{ij}\xi',\,i\in I_{X_j}
\enskip.
\end{equation*}

We need to show that they are simulated by the transitions:
\begin{equation*}
(s_{\suffset{\xi}{0}},\stack{\xi})\step{\alpha_{ij}}(s_{\suffset{\xi_{ij}\xi'}{0}},\stack{\xi_{ij}\xi'}),\,i\in I_{X_j}
\enskip.
\end{equation*}
Thus we have $(x_{ij},(s_{\suffset{\xi_{ij}\xi'}{0}},\stack{\xi_{ij}\xi'}))\in\R$.

We consider the configuration $(s_{\suffset{\xi}{0}},\stack{\xi})$, we distinguish two cases of the top symbol of the stack.
\begin{enumerate}
\item If $\get{\stack{\xi}}{1}=X^{\dagger}_{j}$, then $\M$ has the transition
\begin{equation*}
(s_{\suffset{\xi}{0}},X^{\dagger}_{j},\alpha_{ij},\delta(s_{\suffset{\xi}{0}},X^{\dagger}_{j},\xi_{ij}),s_{D(s_{\suffset{\xi}{0}},X^{\dagger}_{j},\xi_{ij})})
\enskip.
\end{equation*}
The new stack is $S=\delta(s_{\suffset{\xi}{0}},X^{\dagger}_{j},\xi_{ij})\stack{\xi'}$. We verify that $S=\stack({\xi_{ij}\xi'})$. Note that they share the same suffix $\stack{\xi'}$. We only needs to verify the first $\length{\xi_{ij}}$ elements. For the $l$-th element, we let $X_l=\get{\xi{ij}}{l}$, and we distinguish with two cases.
\begin{enumerate}
\item If $X_l\notin(\suffset{\xi}{0}/\{X_j\})\cup \suffset{\xi_{ij}}{l}$, then the $l$-th element of $S$ is $X^{\dagger}_{l}$. Since $\get{\stack{\xi}}{1}=X^{\dagger}_{j}$, from the definition of $\mathit{stack}$, we have $X_j\notin\suffset{\xi}{1}=\suffset{\xi'}{0}$. Therefore, $\suffset{\xi}{0}/\{X_j\}=\suffset{\xi'}{0}$. In this case, $X_l\notin\suffset{\xi'}{0}\cup\suffset{\xi_{ij}}{l}$. Moreover, we have $X_l\notin\suffset{\xi_{ij}\xi'}{l}$, therefore, the $l$-th element of $\stack{\xi_{ij}\xi'}$ is also $X^{\dagger}_l$.
\item Otherwise, then the $l$-th element of $S$ is $X_{l}$. By the definition of $\mathit{stack}$, we get that the $l$-th element of $\stack{\xi_{ij}\xi'}$ is also $X_l$.
\end{enumerate}
Moreover, we verify that the new state $s_{D(s_{\suffset{\xi}{0}},X^{\dagger}_{j},\xi_{ij})}=s_{\suffset{\xi_{ij}\xi'}{0}}$.
Note that we have
\begin{eqnarray*}
&&D(s_{\suffset{\xi}{0}},X^{\dagger}_{j},\xi_{ij})=(\suffset{\xi}{0}/\{X_j\})\cup\suffset{\xi_{ij}}{0}\\
&&=\suffset{\xi'}{0}\cup\suffset{\xi_{ij}}{0}=\suffset{\xi_{ij}\xi'}{0}
\enskip.
\end{eqnarray*}

Hence, we have $(s_{\suffset{\xi}{0}},\stack{\xi})\step{\alpha_{ij}}(s_{\suffset{\xi_{ij}\xi'}{0}},\stack{\xi_{ij}\xi'})$.
\item if  $\get{\stack{\xi}}{1}=X_{j}$, then $\M$ has the transition
\begin{equation*}
(s_{\suffset{\xi}{0}},X_{j},\alpha_{ij},\delta(s_{\suffset{\xi}{0}},X_{j},\xi_{ij}),s_{D(s_{\suffset{\xi}{0}},X_{j},\xi_{ij})})
\enskip,
\end{equation*}
The ne stack is $S=\delta(s_{\suffset{\xi}{0}},X_{j},\xi_{ij})\stack{\xi'}$. We verify that $S=\stack({\xi_{ij}\xi'})$. Note that they share the same suffix $\stack{\xi'}$. We only needs to verify the first $\length{\xi_{ij}}$ elements. For the $l$-th element, we let $X_l=\get{\xi{ij}}{l}$, and we distinguish with two cases.
\begin{enumerate}
\item If $X_l\notin(\suffset{\xi}{0})\cup \suffset{\xi_{ij}}{l}$, then the $l$-th element of $S$ is $X^{\dagger}_{l}$. Since $\get{\stack{\xi}}{1}=X_{j}$, from the definition of $\mathit{stack}$, we have $X_j\in\suffset{\xi}{1}=\suffset{\xi'}{0}$. Therefore, $\suffset{\xi}{0}=\suffset{\xi'}{0}$. In this case, $X_l\notin\suffset{\xi'}{0}\cup\suffset{\xi_{ij}}{l}$. Moreover, we have $X_l\notin\suffset{\xi_{ij}\xi'}{l}$, therefore, the $l$-th element of $\stack{\xi_{ij}\xi'}$ is also $X^{\dagger}_l$.
\item Otherwise, then the $l$-th element of $S$ is $X_{l}$. By the definition of $\mathit{stack}$, we get that the $l$-th element of $\stack{\xi_{ij}\xi'}$ is also $X_l$.
\end{enumerate}
Moreover, we verify that the new state $s_{D(s_{\suffset{\xi}{0}},X_{j},\xi_{ij})}=s_{\suffset{\xi_{ij}\xi'}{0}}$.
Note that we have
\begin{eqnarray*}
&&D(s_{\suffset{\xi}{0}},X_{j},\xi_{ij})=\suffset{\xi}{0}\cup\suffset{\xi_{ij}}{0}\\
&&=\suffset{\xi'}{0}\cup\suffset{\xi_{ij}}{0}=\suffset{\xi_{ij}\xi'}{0}
\enskip.
\end{eqnarray*}
Hence, we have $(s_{\suffset{\xi}{0}},\stack{\xi})\step{\alpha_{ij}}(s_{\suffset{\xi_{ij}\xi'}{0}},\stack{\xi_{ij}\xi'})$.
\end{enumerate}
By concluding the two cases, the above transitions are correct.

Using a similar analysis, we also have all the transitions from $(s_{\suffset{\xi}{0}},\stack{\xi})$ are simulated by $X_j\xi'$.

Now we consider the termination condition. $\xi\downarrow$ iff for all $X\in\suffset{\xi}{0}$, $X\downarrow$. Note that $(s_{\suffset{\xi}{0}},\stack{\xi})\downarrow$ iff for all $X\in\suffset{\xi}{0}$, $X\downarrow$. Therefore, termination condition is also verified.

Hence, we have $\T(X_0)\bisim \T(\M)$.
\end{proof}
}

We have the following theorem.
\begin{theorem}~\label{thm:cfp-pda}
For every context-free process $P$, there exists a \PDA{} $\M$, such that $\T(P)\bisim \T(\M)$.
\end{theorem}

\section{Executability in the Context of Termination}\label{sec:Termination}
In this section, we shall discuss the theory of executability in the context of termination. We shall prove that \TCPN{} is reactively Turing powerful in the context of termination.

The notion of reactive Turing machines (RTM)~\cite{BLT2013} was introduced as an extension of Turing machines to define which behaviour is executable by a computing system in terms of labelled transition systems.
 The definition of RTMs is parameterised with the set $\Atau$, which we assume to be a finite set. Furthermore, the definition is parameterised with another finite set $\D$ of \emph{data symbols}. We extend $\D$ with a special symbol $\Box\notin\D$ to denote a blank tape cell, and denote the set $\D\cup\{\Box\}$ of \emph{tape symbols} by $\Dbox$.
\begin{definition}
[Reactive Turing Machine]\label{def:rtm}
A \emph{reactive Turing machine} (RTM) is a quadruple $(\Sta,\step{},\uparrow,\downarrow)$, where
\begin{enumerate}
    \item $\Sta$ is a finite set of \emph{states},
    \item ${\step{}}\subseteq \Sta\times\Dbox\times\Atau\times\Dbox\times\{L,R\}\times\Sta$ is a finite collection of $(\Dbox\times\Atau\times\Dbox\times\{L,R\})$-labelled \emph{transitions} (we write $s\step{a[d/e]M}t$ for $(s,d,a,e,M,t)\in{\step{}}$),
    \item ${\uparrow}\in\Sta$ is a distinguished \emph{initial state}.
    \item ${\downarrow}\subseteq\Sta$ is a finite set of \emph{final states}.
\end{enumerate}
\end{definition}

Intuitively, the meaning of  a transition $s\step{a[d/e]M}t$ is that whenever the RTM is in state $s$, and $d$ is the symbol currently read by the tape head, then it may execute the action $a$, write symbol $e$ on the tape (replacing $d$), move the read/write head one position to the left or the right on the tape (depending on whether $M=L$ or $M=R$), and then end up in state $t$.

To formalise the intuitive understanding of the operational behaviour of RTMs, we associate with every RTM $\M$ an $\Atau$-labelled transition system  $\T(\M)$. The states of $\T(\M)$ are the configurations of $\M$, which consist of a state from $\Sta$, its tape contents, and the position of the read/write head.
We denote by $\check{\Dbox}=\{\check{d}\mid d\in\Dbox\}$ the set of \emph{marked} symbols; a \emph{tape instance} is a sequence $\delta\in(\Dbox\cup\check{\Dbox})^{*}$ such that $\delta$ contains exactly one element of the set of marked symbols $\check{\Dbox}$, indicating the position of the read/write head.
We adopt a convention to concisely denote an update of the placement of the tape head marker. Let $\delta$ be an element of $\Dbox^{*}$. Then by $\tphdL{\delta}$ we denote the element of $(\Dbox\cup\check{\Dbox})^{*}$ obtained by placing the tape head marker on the right-most symbol of $\delta$ (if it exists), and $\check{\Box}$ otherwise.
Similarly $\tphdR{\delta}$ is obtained by placing the tape head marker on the left-most symbol of $\delta$ (if it exists), and $\check{\Box}$ otherwise.

\begin{definition}\label{def:lts-tm}
Let $\M=(\Sta,\step{},\uparrow,\downarrow)$ be an RTM. The \emph{transition system} $\T(\M)$ \emph{associated with} $\M$ is defined as follows:
\begin{enumerate}
\item its set of states is the set $\Conf[\M]=\{(s,\delta)\mid s\in\Sta,\ \text{$\delta$ a tape instance}\}$ of all configurations of $\M$;
    \item its transition relation ${\step{}}\subseteq{\Conf[\M]\times\Atau\times\Conf[\M]}$ is a relation satisfying, for all $a\in\Atau,\,d,e\in\Dbox$ and $\delta_L,\delta_R\in\Dbox^{*}$:
    \begin{itemize}
        \item $(s,\delta_L\check{d}\delta_R)\step{a}(t,\tphdL{\delta_L}e\delta_R)$ iff $s\step{a[d/e]L}t$,
        \item $(s,\delta_L\check{d}\delta_R)\step{a}(t,\delta_L e{}\tphdR{\delta_R})$ iff $s\step{a[d/e]R}t$;
    \end{itemize}
    \item its initial state is the configuration $(\uparrow,\check{\Box})$; and
    \item its set of final states is the set $\{(s,\delta)\mid \text{$\delta$ a tape instance},\, s\downarrow\}$.
\end{enumerate}
\end{definition}

Turing introduced his machines to define the notion of \emph{effectively computable function} in~\cite{Turing1936}. By analogy, the notion of RTM can be used to define a notion of \emph{effectively executable behaviour}.

\begin{definition}
[Executability]\label{def:exe}
A transition system is \emph{executable} if it is the transition system associated with some RTM.
\end{definition}

In the theory of executability, we use the notion of executable transition systems to evaluate the absolute expressiveness of process calculi in two aspects. On the one hand, if very transition system associated with a process expression specified in a process calculus is executable modulo some behavioural equivalence, then we say that the process calculus is \emph{executable} modulo that behavioural equivalence. One the other hand, if every executable transition system is behavioural equivalent to some transition system associated with a process expression specified in a process calculus modulo some behavioural equivalence, then we say that the process calculus is \emph{reactively Turing powerful} modulo that behavioural equivalence.

We only brief explain that both \TCPS{} and \TCPN{} are executable. We observe that their transition systems are effective and does not contain any unbounded branching behaviour. Thus we can apply the result from~\cite{BLT2013} and draw a conclusion that they are executable modulo $\bbisimd$.

Now we emphasis on showing that \TCPN{} is a reactively Turing powerful process calculus modulo $\bbisimd$.

 We first introduce the notion of bisimulation up to $\bbisim$, which is a useful tool to establish the proofs in this section. Note that we adopt a non-symmetric bisimulation up to relation.

\begin{definition}\label{def:up-to}
Let $T=(\Sta,\step{},\uparrow,\downarrow)$ a transition system. A relation $\R\subseteq\Sta\times\Sta$ is a bisimulation up to $\bbisim$ if, whenever $s_1\R s_2$, then for all $a\in \Atau$:
\begin{enumerate}
    \item if $s_1\step{}^{*}s_1''\step{a}s_1'$, with $s_1\bbisim s_1''$ and ${a\neq\tau}\vee{s_1''\not\bbisim s_1'}$, then there exists $s_2'$ such that $s_2\step{a}s_2'$, $s_1''\mathrel{\bbisim\mathrel{\circ}\mathrel{\R}}s_2$ and $s_1'\mathrel{\bbisim \mathrel{\circ} \mathrel{\R}} s_2'$;
    \item if $s_2\step{a}s_2'$, then there exist $s_1',s_1''$ such that $s_1\step{}^{*}s_1''\step{a}s_1'$, $s_1''\bbisim s_1$ and $s_1'\mathrel{\bbisim \mathrel{\circ} \mathrel{\R}} s_2'$;
    \item if $s_1\downarrow$, then there exists $s_2'$ such that $s_2\step{}^{*} s_2'$ and $s_2'\downarrow$; and
    \item if $s_2\downarrow$, then there exists $s_1'$ such that $s_1\step{}^{*} s_1'$ and $s_1'\downarrow$;
\end{enumerate}
\end{definition}

\begin{lemma}\label{lemma:up-to}
If $\R$ is a bisimulation up to $\bbisim$, then $\R \subseteq {\bbisim}$.
\end{lemma}
\arx{
\begin{proof}
It is sufficient to prove that $\mathrel{\bbisim \mathrel{\circ} \mathrel{\R}}$ is a branching bisimulation, for $\bbisim$ is an equivalence relation.
Let $s_1,s_2,s_3\in\Sta$ and $s_1\bbisim s_2\mathrel{\R} s_3$.
\begin{enumerate}
    \item Suppose $s_1\step{a}s_1'$. We distinguish two cases:
    \begin{enumerate}
        \item If $a=\tau$ and $s_1\bbisim s_1'$, then $s_1'\bbisim s_1\bbisim s_2$, so $s_1'\mathrel{\bbisim \mathrel{\circ} \mathrel{\R}}s_3$. It satisfies Condition 1 of the definition of branching bisimulation.
        \item Otherwise, we have ${a\neq\tau}\vee{s_1\not\bbisim s_1'}$. Then, since $s_1 \bbisim s_2$, according to Definition~\ref{def:bbisim}, there exist $s_2''$ and $s_2'$ such that $s_2\step{}^{*}s_2''\step{a}s_2'$, $s_1\bbisim s_2''$ and $s_1'\bbisim s_2'$. Note that $s_2\bbisim s_1 \bbisim s_2''$, and it is needed to apply Condition 1 of Definition~\ref{def:up-to}. Then we have there exist $s_4''$, $s_4'$ and $s_3'$ such that $s_3\step{a}s_3'$ and $s_2''\bbisim s_4'' \mathrel{\R} s_3$ and $s_2'\bbisim s_4'\mathrel{\R} s_3'$. Since $s_1'\bbisim s_2'\bbisim s_4'$ and $s_4' \mathrel{\R} s_3' $, it follows that $s_1'\mathrel{\bbisim \mathrel{\circ} \mathrel{\R}} s_3'$. It satisfies Condition 1 of the definition of branching bisimulation.

    \end{enumerate}
    \item If $s_3\step{a}s_3'$, then according to Definition~\ref{def:up-to}, there exist $s_2''$ and $s_2'$ such that $s_2\step{}^{*}s_2''\step{a}s_2'$, $s_2''\bbisim s_2$ and $s_2'\mathrel{\bbisim \mathrel{\circ} \mathrel{\R}}s_3'$ , since $s_1\bbisim s_2\bbisim s_2''$ and $s_2''\step{a}s_2'$, by Definition~\ref{def:bbisim}, there exist $s_1''$ and $s_1'$ such that $s_1\step{}^{*}s_1''\step{(a)}s_1'$ with $s_1'' \bbisim s_2''$ and $s_1'\bbisim s_2'$. Since $s_2''\mathrel{\bbisim \mathrel{\circ}\mathrel{\R}} s_3$ and $s_2'\mathrel{\bbisim \mathrel{\circ} \mathrel{\R}}s_3'$, it follows that $s_1''\mathrel{\bbisim \mathrel{\circ} \mathrel{\R}}s_3$ and $s_1'\mathrel{\bbisim \mathrel{\circ} \mathrel{\R}}s_3'$. It satisfies the symmetry of Condition 1 of the definition of branching bisimulation.
\end{enumerate}
The termination condition is also satisfied from Definition~\ref{def:up-to}.

Therefore, a branching bisimulation up to $\bbisim$ is included in $\bbisim$.
\end{proof}
}

Next we show that \TCPN{} is a reactively Turing powerful by writing a specification of the reactive Turing machine with \TCPN modulo $\bbisimd$. The proof consists of five steps.
\begin{enumerate}
\item We first write the specification of a terminating half counter;
\item then we show that every regular process can be specified in \TCPN;
\item next we use two half counters and a regular process to encode a terminating stack;
\item with two stacks and a regular process we can specify a tape; and
\item finally we use a tape and a regular control process to specify an RTM.
\end{enumerate}

We first recall the following infinite specification in \SAS{} of a terminating half counter:
\begin{eqnarray*}
C_n &=& \mathit{a}.C_{n+1}+\mathit{b}.B_{n}+\one\,(n\in\mathbb{N})\\
B_n &=& \mathit{a}.B_{n-1}+\one\,(n\geq 1)\\
B_0 &=& \mathit{c}.C_0+\one
\end{eqnarray*}

We provide a specification of a counter in \TCPN{} as follows:
\begin{equation*}
HC=(\nest{\mathit{(a+\one)}}{\mathit{(b+\one)}}\bullet (c+\one))^{*}
\end{equation*}

We have the following lemma:

\begin{lemma}~\label{lemma:tcpn-halfcounter}
$C_0\bbisimd HC$
\end{lemma}
\arx{
\begin{proof}
We verify that $HC\bbisimd C_0$. Consider the following relation:
\begin{equation*}
\R_1=\{(C_0, HC)\}\cup\{(C_n,\nest{(a+\one)}{(b+\one)}\bullet (a+\one)^n\bullet (c+\one)\bullet HC)\mid n\geq 1\}\cup\{(B_n,(a+\one)^n\bullet (c+\one)\bullet HC)\mid n\in\mathbb{N}\}
\enskip.
\end{equation*}

We let $\R_2$ be the symmetrical relation of $\R_1$. We show that $\R=\R_1\cup\R_2$ is a divergence-preserving branching bisimulation as follows:

Note that $\R$ satisfies the divergence-preserving condition since there is no infinite sequence of $\tau$ transitions.
In this prove, we only illustrate the pairs in $\R_1$, since we can use the symmetrical argument for the pairs in $\R_2$. We first consider the pair $(C_0,HC)$. Note that $C_0$ has the following transitions:
\begin{eqnarray*}
&C_0\step{a} C_1,\,\mbox{and}\\
&C_0\step{b} B_0
\enskip,
\end{eqnarray*}
which are simulated by:
\begin{eqnarray*}
&HC\step{a}\nest{(a+\one)}{(b+\one)}\bullet (a+\one)\bullet (c+\one)\bullet HC,\, \mbox{and}\\
&HC\step{b}(c+\one)\bullet HC
\enskip,
\end{eqnarray*}
with $(C_1,\nest{(a+\one)}{(b+\one)}\bullet (a+\one)\bullet (c+\one)\bullet HC)\in\R$ and $(B_0,(c+\one)\bullet HC)\in\R$. Moreover, we have $C_0\downarrow$ and $HC\downarrow$.

Now we consider the pair $(C_n,\nest{(a+\one)}{(b+\one)}\bullet (a+\one)^n\bullet (c+\one)\bullet HC)$, with $n\geq 1$. Note that $C_n$ has the following transitions:
\begin{eqnarray*}
&C_n\step{a}C_{n+1},\,\mbox{and}\\
&C_n\step{b}B_{n}
\enskip,
\end{eqnarray*}
which are simulated by:
\begin{eqnarray*}
&\nest{(a+\one)}{(b+\one)}\bullet (a+\one)^n\bullet (c+\one)\bullet HC\step{a} \nest{(a+\one)}{(b+\one)}\bullet (a+\one)^{n+1}\bullet (c+\one)\bullet HC,\,\mbox{and}\\
&\nest{(a+\one)}{(b+\one)}\bullet (a+\one)^n\bullet (c+\one)\bullet HC\step{b}(a+\one)^n\bullet (c+\one)\bullet HC
\enskip,
\end{eqnarray*}
with $(C_{n+1},\nest{(a+\one)}{(b+\one)}\bullet (a+\one)^{n+1}\bullet (c+\one)\bullet HC)\in\R$ and $(B_n,(a+\one)^n\bullet (c+\one)\bullet HC)\in\R$. Moreover, we have $C_n\downarrow$ and $\nest{(a+\one)}{(b+\one)}\bullet (a+\one)^n\bullet (c+\one)\bullet HC\downarrow$.

Now we proceed to consider the pair $(B_0,(c+\one)\bullet HC)$. Note that $B_0$ has the following transition:
\begin{eqnarray*}
&B_0\step{c}C_0
\enskip,
\end{eqnarray*}
which is simulated by:
\begin{eqnarray*}
&(c+\one)\bullet HC\step{c}HC
\enskip,
\end{eqnarray*}
with $(C_0,HC)\in\R$. Moreover, we have $B_0\downarrow$ and $(c+\one)\bullet HC\downarrow$.

Next we consider the pair $(B_n,(a+\one)^n\bullet (c+\one)\bullet HC)$, with $n\geq 1$. Note that $B_n$ has the following transition:
\begin{eqnarray*}
&B_n\step{a}B_{n-1}
\enskip,
\end{eqnarray*}
which is simulated by:
\begin{eqnarray*}
&(a+\one)^n\bullet (c+\one)\bullet HC\step{a}(a+\one)^{n-1}\bullet (c+\one)\bullet HC
\enskip,
\end{eqnarray*}
with $(B_{n-1},(a+\one)^{n-1}\bullet (c+\one)\bullet HC)\in\R$. Moreover, we have $B_n\downarrow$ and $(a+\one)^n\bullet (c+\one)\bullet HC\downarrow$.

Hence, we have $C_0\bbisimd HC$.
\end{proof}
}

Next we show that every regular process can be specified in \TCPN{} modulo $\bbisimd$.
A regular process with at finite set of action labels $\Atau$ is given by $P_i=\sum_{j=1}^{n} \alpha_{ij}\bullet P_j +\beta_i\,(i=1,\ldots,n)$ where $\alpha_{ij}$ and $\beta_{i}$ are finite sums of actions from $\Atau$. We show the following lemma.

\begin{lemma}~\label{lemma:tcpn-rular}
Every regular process can be specified in \TCPN{} modulo $\bbisimd$.
\end{lemma}
\arx{
\begin{proof}
We consider a regular process with at finite set of action labels $\Atau$ which is given by $P_i=\sum_{j=1}^{n} \alpha_{ij}\bullet P_j +\beta_i\,(i=1,\ldots,n)$ where $\alpha_{ij}$ and $\beta_{i}$ are finite sums of actions from $\Atau$.  We let $c!0,c!1,\ldots,c!(n+1),c?0,c?1,\ldots,c?(n+1)$ be labels that are not in $\Atau$.

Consider the following process:
\begin{eqnarray*}
G_i&=&\sum_{j=1}^{n}\alpha_{ij}\bullet (c!j+\one) +\beta_i\bullet (c!0+\one)\\
M&=&\nest{\left(\sum_{j=1}^{n}(c?j+\one)\bullet G_j+(c!(n+1)+\one)\bullet(c?(n+1)+\one)\right)}{(c?0+\one)}\\
N&=&\nest{\left(\sum_{j=1}^{n+1}(c?j+\one)\bullet(c!j+\one)\right)}{((c?0+\one)\bullet(c!0+\one))}
\end{eqnarray*}

Note that $\bullet$ is associative and we suppose that $\bullet$ binds stronger than $+$.
We verify that $P_i\bbisimd [G_i\bullet M\parallel N]_{\{c\}}$.
We let $Q=\left(\sum_{j=1}^{n}(c?j+\one)\bullet G_j+(c!(n+1)+\one)\bullet (c?(n+1)+\one)\right)$ and $O=\left(\sum_{j=1}^{n+1}(c?j+\one)\bullet(c!j+\one)\right)$. We let
\begin{eqnarray*}
\R_1&=&\{(P_i, [G_i\bullet M\bullet Q^{k}\parallel N\bullet O^{k}]_{\{c\}})\mid k\in\mathbb{N},\,i=1,\ldots,n\}\\
&\cup&\{(P_i,[(c!i+\one)\bullet M\bullet Q^{k}\parallel N\bullet O^{k}]_{\{c\}})\mid k\in\mathbb{N},\,i=1,\ldots,n\}\\
&\cup&\{(P_i,[M\bullet Q^{k}\parallel (c!i+\one)\bullet N\bullet O^{k+1}]_{\{c\}})\mid k\in\mathbb{N},\,i=1,\ldots,n\}\\
&\cup&\{(\one,[(c!0+\one)\bullet M\bullet Q^{k}\parallel N\bullet O^{k}]_{\{c\}})\mid k\in\mathbb{N}\}\\
&\cup&\{(\one,[M\bullet Q^{k}\parallel (c!0+\one)\bullet O^{k}]_{\{c\}})\mid k\in\mathbb{N}\}\\
&\cup&\{(\one,[Q^{k}\parallel O^{k}]_{\{c\}})\mid k\in\mathbb{N}\}\\
&\cup&\{(\one,[(c?(n+1)+\one)\bullet Q^{k}\parallel (c!(n+1)+\one)\bullet O^{k}]_{\{c\}})\mid k\in\mathbb{N}\}
\enskip;
\end{eqnarray*}
and we let $\R_2$ be the symmetrical relation of $\R_1$. We show that $\R=\R_1\cup\R_2$ is a divergence-preserving branching bisimulation. We shall only verify the pairs in $\R_1$ in this proof since $\R$ is symmetrical.

For the set of pairs $\{(P_i, [G_i\bullet M\bullet Q^{k}\parallel N\bullet O^{k}]_{\{c\}})\mid k\in\mathbb{N},\,i=1,\ldots,n\}$, note that $P_i$ has the following transitions: $P_i\step{a}P_j$ if $a$ is a summand of $\alpha_{ij}$, or $P_i\step{a}\one$ if $a$ is a summand of $\beta_j$.

The first transition is simulated by the following transitions:
\begin{eqnarray*}
&&[G_i\bullet M\bullet Q^{k}\parallel N\bullet O^{k}]_{\{c\}}\step{a}[(c!j+\one)\bullet M\bullet Q^{k}\parallel N\bullet O^{k}]_{\{c\}}\\
&&\step{\tau}[M\bullet Q^{k}\parallel (c!j+\one) \bullet N\bullet O^{k+1}]_{\{c\}}\\
&&\step{\tau} [G_j\bullet M\bullet Q^{k+1}\parallel N\bullet O^{k+1}]_{\{c\}}
\enskip.
\end{eqnarray*}

If $k\geq 1$, then the second transition is simulated by the following transitions:
\begin{eqnarray*}
&&[G_i\bullet M\bullet Q^{k}\parallel N\bullet O^{k}]_{\{c\}}\step{a}[(c!0+\one)\bullet M\bullet Q^{k}\parallel N\bullet O^{k}]_{\{c\}}\\
&&\step{\tau}[M\bullet Q^{k}\parallel (c!0+\one)\bullet O^{k}]_{\{c\}}\step{\tau}[Q^{k}\parallel O^{k}]_{\{c\}}\\
&&\step{\tau}[(c?(n+1)+\one)\bullet Q^{k-1}\parallel (c!(n+1)+\one)\bullet O^{k-1}]_{\{c\}}\step{\tau}[Q^{k-1}\parallel O^{k-1}]_{\{c\}}\\
&&\step{}^{*}\one
\enskip;
\end{eqnarray*}

otherwise, if $k=0$, then the second transition are simulated by:

\begin{eqnarray*}
&&[G_i\bullet M\parallel N]_{\{c\}}\step{a}[(c!0+\one)\bullet M\parallel N]_{\{c\}}\\
&&\step{\tau}[M\parallel (c!0+\one)]_{\{c\}}\step{\tau}\one
\enskip.
\end{eqnarray*}

We have that that $(P_j,[(c!j+\one)\bullet M\bullet Q^{k}\parallel N\bullet O^{k}]_{\{c\}})\in\R$, $(P_j,[M\bullet Q^{k}\parallel (c!j+\one) \bullet N\bullet O^{k+1}]_{\{c\}})\in\R$, $(P_j,[G_j\bullet M\bullet Q^{k+1}\parallel N\bullet O^{k+1}]_{\{c\}})\in\R$, $(\one,[(c!0+\one)\bullet M\bullet Q^{k}\parallel N\bullet O^{k}]_{\{c\}})\in\R$, $(\one,[M\bullet Q^{k}\parallel (c!0+\one)\bullet O^{k}]_{\{c\}})\in\R$, $(\one,[Q^{k}\parallel O^{k}]_{\{c\}})$, $(\one,[(c?(n+1)+\one)\bullet Q^{k}\parallel (c!(n+1)+\one)\bullet O^{k}]_{\{c\}})\in\R$ and $(\one,\one)\in\R$ for all $k\in\mathbb{N}$ and $i,j=1,\ldots, n$.

One can easily verify that all the other pairs satisfy the condition of branching bisimulation. The relation $\R$ also satisfies the divergence-preserving condition since no infinite $\tau$-transition sequence is allowed from any process defined in $\R$.

Therefore, we get a finite specification of every regular process in \TCPN{} modulo $\bbisimd$.
\end{proof}}

Now we show that a stack can be specified by a regular process and two half counters. We first give an infinite specification in \SAS{} of a stack as follows:
\begin{eqnarray*}
S_{\epsilon}&=&\Sigma_{d\in\Dbox} \mathit{push}?d.S_{d}+\mathit{pop}!\Box.S_{\epsilon}+\one\\
S_{d\delta}&=& \mathit{pop}!d.S_{\delta}+\Sigma_{e\in\Dbox} \mathit{push}?e.S_{ed\delta}+\one
\enskip.
\end{eqnarray*}

Note that $\Dbox$ is a finite set of symbols. We suppose that $\Dbox$ contains $N$ symbols (including $\Box$).
We use $\epsilon$ to denote empty sequence.
We first define an encoding from sequence of symbols to natural numbers $\encode{\_}:{\Dbox}^{*}\Rightarrow\mathbb{N}$ inductively defined as follows:
\begin{eqnarray*}
\encode{\epsilon}&=&0\\
\encode{d_k}&=&k\,(k=1,2,\ldots,N)\\
\encode{d_k\sigma}&=&k+N\times\encode{\sigma}
\enskip.
\end{eqnarray*}
We define a stack in \TCPN{} as follows:
\begin{eqnarray*}
S&=&[X_{\emptyset}\parallel P_1\parallel P_2]_{\{a_1,a_2,b_1,b_2,c_1,c_2\}}\\
P_j&=&(\nest{(a_j!a+\one)}{(b_j!b+\one)}\bullet(c_j!c+\one))^{*}\,(j=1,2)\\
X_{\emptyset}&=&(\Sigma_{j=1}^{N}((\mathit{push}?d_j+\one)\bullet(a_1?a+\one)^{j}\bullet(b_1+\one)\bullet X_j)+\mathit{pop!\Box})^{*}\\
X_{k}&=&\Sigma_{j=1}^{N}((\mathit{push}?d_j+\one)\bullet \mathit{Push_j})+(\mathit{pop}!d_k+\one)\bullet \mathit{Pop_k}\,(k=1,2,\ldots,N)\\
\mathit{Push_k}&=&\mathit{Shift1to2}\bullet(a_1?a+\one)^k\bullet\mathit{NShift2to1}\bullet X_{k}\,(k=1,2,\ldots,N)\\
\mathit{Pop_k}&=&(a_1?a+\one)^k\bullet\mathit{1/NShift1to2}\bullet\mathit{Test_{\emptyset}}\\
\mathit{Shift1to2}&=&((a_1?a+\one)\bullet(a_2?a+\one))^{*}\bullet(c_1?c+\one)\bullet(b_2?b+\one)\\
\mathit{NShift2to1}&=&((a_2?a+\one)\bullet(a_1?a+\one)^{N})^{*}\bullet(c_2?c+\one)\bullet(b_1?b+\one)\\
\mathit{1/NShift1to2}&=&((a_1?a+\one)^{N}\bullet(a_2?a+\one))^{*}\bullet(c_1?c+\one)\bullet(b_2?b+\one)\\
\mathit{Test_{\emptyset}}&=&(a_2?a+\one)\bullet(a_1?a+\one)\bullet\mathit{Test_1}+(c_2?c+\one)\bullet X_{\emptyset}\\
\mathit{Test_{1}}&=&(a_2?a+\one)\bullet(a_1?a+\one)\bullet\mathit{Test_2}+(c_2?c+\one)\bullet X_{1}\\
\mathit{Test_{2}}&=&(a_2?a+\one)\bullet(a_1?a+\one)\bullet\mathit{Test_3}+(c_2?c+\one)\bullet X_{2}\\
&\cdots&\\
\mathit{Test_{N}}&=&(a_2?a+\one)\bullet(a_1?a+\one)\bullet\mathit{Test_1}+(c_2?c+\one)\bullet X_{N}
\enskip.
\end{eqnarray*}

We have the following result.
\begin{lemma}~\label{lemma:tcpn-stack}
$S_{\epsilon}\bbisimd S$.
\end{lemma}
\arx{
\begin{proof}
We define some auxiliary process:
\begin{eqnarray*}
P_j(0)&=&(\nest{(a_j!a+\one)}{(b_j!b+\one)}\bullet(c_j!c+\one))^{*}\,(j=1,2)\\
P_j(n)&=&\nest{(a_j!a+\one)}{(b_j!b+\one)}\bullet (a_j!a+\one)^{n}\bullet(c_j!c+\one))\bullet P_j,\,(j=1,2;\,n=1,2,\ldots)\\
Q_j(n)&=&(a_j!a+\one)^{n}\bullet(c_j!c+\one))\bullet P_j,\,(j=1,2;\,n\in\mathbb{N})
\enskip.
\end{eqnarray*}
$P_0$ and $P_1$ behave as two half counters.

We let $\R_1=\{(S_{\epsilon},S)\}\cup\{(S_{d_j\delta},[X_{j}\bullet X_{\epsilon}\parallel Q_1(m)\parallel P_2(0)]_{\{a_1,a_2,b_1,b_2,c_1,c_2\}})\mid j=\encode{d_j},m=\encode{d_j\delta},d\in\Dbox,\delta\in \Dbox^{*}\}$. We let $\R_2$ be the symmetrical relation of $\R_1$. We verify that $\R=\R_1\cup\R_2\cup\bbisimd$ is a divergence-preserving branching bisimulation relation.

Note that $S_{\epsilon}$ has the following transitions:
\begin{eqnarray*}
&&S_{\epsilon}\step{\mathit{push}?d_j}S_{d_j}\mbox{ for all }j=1,2,\ldots,N,\mbox{ and}\\
&&S_{\epsilon}\step{\mathit{pop}!\Box}S_{\epsilon}
\enskip.
\end{eqnarray*}

They are simulated by the following transitions:
\begin{eqnarray*}
&&S\step{\mathit{push}?d_j}[(a_1?a+\one)^{j}\bullet(b_1+\one)\bullet X_j\bullet X_{\epsilon}\parallel P_1(0)\parallel P_2(0)]_{\{a_1,a_2,b_1,b_2,c_1,c_2\}}\\
&&\step{}^{*}[(b_1+\one)\bullet X_j\bullet X_{\epsilon}\parallel P_1(j)\parallel P_2(0)]_{\{a_1,a_2,b_1,b_2,c_1,c_2\}}\\
&&\step{}^{*}[X_j\bullet X_{\epsilon}\parallel Q_1(j)\parallel P_2(0)]_{\{a_1,a_2,b_1,b_2,c_1,c_2\}}\mbox{ for all }j=1,2,\ldots,N,\mbox{ and}\\
&&S\step{\mathit{pop}!\Box}S
\enskip.
\end{eqnarray*}
We only consider the first case, since the second transition is trivial. We have $(S_{d_j},[X_j\bullet X_{\epsilon}\parallel Q_1(j)\parallel P_2(0)]_{\{a_1,a_2,b_1,b_2,c_1,c_2\}})\in\R$. We denote the sequence of transitions $[(a_1?a+\one)^{j}\bullet(b_1+\one)\bullet X_j\bullet X_{\epsilon}\parallel P_1(0)\parallel P_2(0)]_{\{a_1,a_2,b_1,b_2,c_1,c_2\}}\step{}^{*}[X_j\bullet X_{\epsilon}\parallel Q_1(j)\parallel P_2(0)]_{\{a_1,a_2,b_1,b_2,c_1,c_2\}})\in\R$ by $s_0\step{}^{*}s_m$. It is obvious that $s_0\bbisimd\ldots s_m$. Therefore, $S\step{\mathit{push}?d_j}s_0$, and $s_0\bbisimd s_m$ with $(S_{d_j},s_m)\in\R$.

Note that $S_{d_j\delta}$ has the following transitions:
\begin{eqnarray*}
&&S_{d_j\delta}\step{\mathit{push}?d_k}S_{d_kd_j\delta}\mbox{ for all }k=1,2,\ldots,N,\mbox{ and}\\
&&S_{d_j\delta}\step{\mathit{pop}!d_j}S_{d_k\delta'}\mbox{, where }d_k\delta'=\delta
\enskip.
\end{eqnarray*}

They are simulated by the following transitions:
\begin{eqnarray*}
&&[X_{j}\bullet X_{\epsilon}\parallel Q_1(\encode{d_j\delta})\parallel P_2(0)]_{\{a_1,a_2,b_1,b_2,c_1,c_2\}}\step{\mathit{push}?d_k}[\mathit{Push_k}\bullet X_{\epsilon}\parallel Q_1(\encode{d_j\delta})\parallel P_2(0)]_{\{a_1,a_2,b_1,b_2,c_1,c_2\}}\\
&&\step{}^{*}[(a_1?a+\one)^k\bullet\mathit{NShift2to1}\bullet X_{k}\bullet X_{\epsilon}\parallel P_1(0)\parallel Q_2(\encode{d_j\delta})]_{\{a_1,a_2,b_1,b_2,c_1,c_2\}}\\
&&\step{}^{*}[\mathit{NShift2to1}\bullet X_{k}\bullet X_{\epsilon}\parallel P_1(\encode{d_k})\parallel Q_2(\encode{d_j\delta})]_{\{a_1,a_2,b_1,b_2,c_1,c_2\}}\\
&&\step{}^{*}[X_{k}\bullet X_{\epsilon}\parallel Q_1(\encode{d_kd_j\delta})\parallel P_2(0)]_{\{a_1,a_2,b_1,b_2,c_1,c_2\}}\mbox{ for all }d_j,d_k\in\Dbox,\,\delta\in\Dbox^{*}\mbox{ and}\\
&&[X_{j}\bullet X_{\epsilon}\parallel Q_1(\encode{d_j\delta})\parallel P_2(0)]_{\{a_1,a_2,b_1,b_2,c_1,c_2\}}\step{\mathit{pop}!d_j}[\mathit{Pop_j}\bullet X_{\epsilon}\parallel Q_1(\encode{d_j\delta})\parallel P_2(0)]_{\{a_1,a_2,b_1,b_2,c_1,c_2\}}\\
&&\step{}^{*}[\mathit{1/NShift1to2}\bullet\mathit{Test_{\emptyset}}\bullet X_{\epsilon}\parallel Q_1(\encode{d_j\delta}-k)\parallel P_2(0)]_{\{a_1,a_2,b_1,b_2,c_1,c_2\}}\\
&&\step{}^{*}[\mathit{Test_{\emptyset}}\bullet X_{\epsilon}\parallel P_1(0)\parallel Q_2(\encode{\delta})]_{\{a_1,a_2,b_1,b_2,c_1,c_2\}}\\
&&\step{}^{*}[X_k\bullet X_{\epsilon}\parallel Q_1(\encode{d_k\delta'}\parallel P_2(0)]_{\{a_1,a_2,b_1,b_2,c_1,c_2\}}\mbox{ for all }d_j\in\Dbox,\,\delta\in\Dbox^{*}\mbox{ and}\,\delta=d_k\delta'
\enskip.
\end{eqnarray*}

We have $(S_{d_kd_j\delta},[X_{k}\bullet X_{\epsilon}\parallel Q_1(\encode{d_kd_j\delta})\parallel P_2(0)]_{\{a_1,a_2,b_1,b_2,c_1,c_2\}})\in\R$ and $(S_{d_k\delta'},[X_k\bullet X_{\epsilon}\parallel Q_1(\encode{d_k\delta'}\parallel P_2(0)]_{\{a_1,a_2,b_1,b_2,c_1,c_2\}})\in\R$. By using a similar analysis with the previous case, we conclude that $\R$ is a bisimulation up to $\bbisim$. By Lemma~\ref{lemma:up-to}, we have $\R\subseteq\bbisim$. Moreover, there is no infinite $\tau$-transition sequence from any process defined above. Therefore, $\R\subseteq\bbisimd$.

Hence, we have $S_{\epsilon}\bbisimd S$.
\end{proof}}

Next we proceed to define the tape by means of two stacks. We consider the following infinite specification in \SAS{} of a tape:
\begin{equation*}
T_{\delta_L\tphd{d}\delta_R}=r!d.T_{\delta_L\tphd{d}\delta_R}+\Sigma_{e\in \Dbox}w?e.T_{\delta_L\tphd{e}\delta_R}+L?m. T_{\tphdL{\delta_L}d\delta_R}+R?m.T_{\delta_Ld\tphdR{\delta_R}}+\one
\enskip.
\end{equation*}

We define the tape process in \TCPN{} as follows:
\begin{eqnarray*}
T&=&[T_{\Box}\parallel S_1\parallel S_2]_{\{\mathit{push_1,pop_1,push_2,pop_2}\}}\\
T_{d}&=&r!d.T_{d}+\Sigma_{e\in\Dbox}w?e.T_{e}+L?m.\mathit{Left_d}+R?m.\mathit{Right_d}+\one\, (d\in\Dbox)\\
\mathit{Left_d}&=&\Sigma_{e\in\Dbox}((\mathit{pop_1?e}+\one)\bullet(\mathit{push_2!d}+\one)\bullet T_{e})\\
\mathit{Right_d}&=&\Sigma_{e\in\Dbox}((\mathit{pop_2?e}+\one)\bullet(\mathit{push_1!d}+\one)\bullet T_{e})
\enskip,
\end{eqnarray*}
where $S_1$ and $S_2$ are two stacks with $\mathit{push_1,pop_1,push_2}$ and $\mathit{pop_2}$ as their interfaces.

We establish the following result.
\begin{lemma}~\label{lemma:tcpn-tape}
$T_{\tphd{\Box}}\bbisimd T$.
\end{lemma}
\arx{
\begin{proof}
We define the following auxiliary processes:
\begin{eqnarray*}
S_1(\delta)&=&[X_{1,k}\parallel Q_1(\encode{\delta})\parallel P_2(0)]_{\{a_1,a_2,b_1,b_2,c_1,c_2\}}\\
S_2(\delta)&=&[X_{2,k}\parallel Q_1(\encode{\delta})\parallel P_2(0)]_{\{a_1,a_2,b_1,b_2,c_1,c_2\}},\,\mbox{where}\,\delta=d_k\delta'
\enskip.
\end{eqnarray*}
$X_{1,k}$ and $X_{2,k}$ is obtained by renaming $\mathit{push}$ and $\mathit{pop}$ in $X_k$ to $\mathit{push_1,\,pop_1,\,push_2}$ and $\mathit{pop_2}$ respectively. We use $\overline{\delta}$ to denote the reverse sequence of $\delta$.

We verify that
\begin{equation*}
\R=\{(T_{\delta_L\tphd{d}\delta_R},[T_{d}\parallel S_1(\overline{\delta_L})\parallel S_2(\delta_R)]_{\{\mathit{push_1,pop_1,push_2,pop_2}\}})\mid d\in\Dbox,\,\delta_L,\delta_R\in\Dbox^{*}\}\subseteq\bbisimd
\enskip.
\end{equation*}

$T_{\delta_L\tphd{d}\delta_R}$ has the following transitions:
\begin{eqnarray*}
&&T_{\delta_L\tphd{d}\delta_R}\step{r!d}T_{\delta_L\tphd{d}\delta_R}\\
&&T_{\delta_L\tphd{d}\delta_R}\step{w?e}T_{\delta_L\tphd{e}\delta_R}\,\mbox{for all}\,e\in\Dbox\\
&&T_{\delta_L\tphd{d}\delta_R}\step{L?m}T_{\tphdL{\delta_L}d\delta_R}\,\mbox{if}\,\delta_L\neq\epsilon\\
&&T_{\delta_L\tphd{d}\delta_R}\step{R?m}T_{\delta_Ld\tphdR{\delta_R}}\,\mbox{if}\,\delta_R\neq\epsilon\\
&&T_{\delta_L\tphd{d}\delta_R}\step{L?m}T_{\epsilon\tphd{\Box}d\delta_R}\,\mbox{if}\,\delta_L=\epsilon\,\mbox{and}\\
&&T_{\delta_L\tphd{d}\delta_R}\step{R?m}T_{\delta_Ld\tphd{\Box}\epsilon}\,\mbox{if}\,\delta_R=\epsilon
\enskip.
\end{eqnarray*}

They are simulated by the following transitions:
\begin{eqnarray*}
&&[T_{d}\parallel S_1(\overline{\delta_L})\parallel S_2(\delta_R)]_{\{\mathit{push_1,pop_1,push_2,pop_2}\}}\step{r!d}[T_{d}\parallel S_1(\overline{\delta_L})\parallel S_2(\delta_R)]_{\{\mathit{push_1,pop_1,push_2,pop_2}\}}\\
&&[T_{d}\parallel S_1(\overline{\delta_L})\parallel S_2(\delta_R)]_{\{\mathit{push_1,pop_1,push_2,pop_2}\}}\step{e?d}[T_{e}\parallel S_1(\overline{\delta_L})\parallel S_2(\delta_R)]_{\{\mathit{push_1,pop_1,push_2,pop_2}\}}\,\mbox{for all}\,e\in\Dbox\\
&&[T_{d}\parallel S_1(\overline{\delta_L})\parallel S_2(\delta_R)]_{\{\mathit{push_1,pop_1,push_2,pop_2}\}}\step{L?m}[\mathit{Left_d}\parallel S_1(\overline{\delta_L})\parallel S_2(\delta_R)]_{\{\mathit{push_1,pop_1,push_2,pop_2}\}}\\
&&\step{}^{*}[T_{e}\parallel S_1(\overline{\delta_L'})\parallel S_2(d\delta_R)]_{\{\mathit{push_1,pop_1,push_2,pop_2}\}},\,\delta_L=\delta_L'e,\,\mbox{if}\,\delta_L\neq\epsilon\\
&&[T_{d}\parallel S_1(\overline{\delta_L})\parallel S_2(\delta_R)]_{\{\mathit{push_1,pop_1,push_2,pop_2}\}}\step{R?m}[\mathit{Right_d}\parallel S_1(\overline{\delta_L})\parallel S_2(\delta_R)]_{\{\mathit{push_1,pop_1,push_2,pop_2}\}}\\
&&\step{}^{*}[T_{e}\parallel S_1(\overline{\delta_Ld})\parallel S_2(\delta_R')]_{\{\mathit{push_1,pop_1,push_2,pop_2}\}},\,\delta_R=e\delta_R,\,\mbox{if}\,\delta_R\neq\epsilon\\
&&[T_{d}\parallel S_1(\overline{\delta_L})\parallel S_2(\delta_R)]_{\{\mathit{push_1,pop_1,push_2,pop_2}\}}\step{L?m}[\mathit{Left_d}\parallel S_1(\overline{\delta_L})\parallel S_2(\delta_R)]_{\{\mathit{push_1,pop_1,push_2,pop_2}\}}\\
&&\step{}^{*}[T_{\Box}\parallel S_1(\epsilon)\parallel S_2(d\delta_R)]_{\{\mathit{push_1,pop_1,push_2,pop_2}\}},\,\mbox{if}\,\delta_L=\epsilon\\
&&[T_{d}\parallel S_1(\overline{\delta_L})\parallel S_2(\delta_R)]_{\{\mathit{push_1,pop_1,push_2,pop_2}\}}\step{R?m}[\mathit{Right_d}\parallel S_1(\overline{\delta_L})\parallel S_2(\delta_R)]_{\{\mathit{push_1,pop_1,push_2,pop_2}\}}\\
&&\step{}^{*}[T_{\Box}\parallel S_1(\overline{\delta_Ld})\parallel S_2(\epsilon)]_{\{\mathit{push_1,pop_1,push_2,pop_2}\}},\,\mbox{if}\,\delta_R=\epsilon
\enskip.
\end{eqnarray*}

We have
\begin{eqnarray*}
&&(T_{\delta_L\tphd{d}\delta_R},[T_{d}\parallel S_1(\overline{\delta_L})\parallel S_2(\delta_R)]_{\{\mathit{push_1,pop_1,push_2,pop_2}\}})\in\R,\\ &&(T_{\delta_L\tphd{e}\delta_R},[T_{e}\parallel S_1(\overline{\delta_L})\parallel S_2(\delta_R)]_{\{\mathit{push_1,pop_1,push_2,pop_2}\}})\in\R,\\ &&(T_{\tphdL{\delta_L}d\delta_R},[T_{e}\parallel S_1(\overline{\delta_L'})\parallel S_2(d\delta_R)]_{\{\mathit{push_1,pop_1,push_2,pop_2}\}})\in\R,\\ &&(T_{\delta_Ld\tphdR{\delta_R}},[T_{e}\parallel S_1(\overline{\delta_Ld})\parallel S_2(\delta_R')]_{\{\mathit{push_1,pop_1,push_2,pop_2}\}})\in\R,\\
&&(T_{\epsilon\tphd{\Box}d\delta_R},[T_{\Box}\parallel S_1(\epsilon)\parallel S_2(d\delta_R)]_{\{\mathit{push_1,pop_1,push_2,pop_2}\}})\in\R,\,\mbox{and}\\
&&(T_{\delta_Ld\tphd{\Box}\epsilon},[T_{\Box}\parallel S_1(\overline{\delta_Ld})\parallel S_2(\epsilon)]_{\{\mathit{push_1,pop_1,push_2,pop_2}\}})\in\R
\enskip.
\end{eqnarray*}

By an analysis similar from Lemma~\ref{lemma:tcpn-stack}, we have $\R$ is a bisimulation up to $\bbisim$. Therefore, $\R\subset\bbisim$. Moreover, there is no infinite $\tau$-transition sequence from the processes defined above. Therefore, $\R\subseteq\bbisimd$.

Hence, we have $T_{\tphd{\Box}}\bbisimd T$.
\end{proof}}

Finally, we construct a finite control process for an RTM $\M=(\Sta_{\M},\step{}_{\M},\uparrow_{\M},\downarrow_{\M})$ as follows:
\begin{equation*}
C_{s,d}=\Sigma_{(s,d,a,e,M,t)\in\step{}_{\M}}(a.w!e.M!m.\Sigma_{f\in\Dbox}r?f.C_{t,f})[+\one]_{s\downarrow_{\M}}\,(s\in\Sta_{\M},d\in\Dbox)
\enskip.
\end{equation*}

We prove the following lemma.
\begin{lemma}~\label{lemma:tcpn-control}
$\T(\M)\bbisimd [C_{\uparrow_{\M},\Box}\parallel T]_{\{r,w,L,R\}}$.
\end{lemma}
\arx{
\begin{proof}
By the proof of Theorem~\ref{thm:congruence}, $\bbisimd$ is compatible with parallel composition. Therefore, it is enough to show that $\T(\M)\bbisimd [C_{\uparrow,\Box}\parallel T_{\tphd{\Box}}]_{\{r,w,L,R\}}$.

We define a binary relation $\R$ by:

\begin{eqnarray*}
\R&=&\{((s,\delta_L\tphd{d}\delta_R),[C_{s,d}\parallel T_{\delta_L\tphd{d}\delta_R}]_{\{r,w,L,R\}})\mid s\in\Sta_M,\,\delta_L,\delta_R\in\Dbox^{*},\,d\in\Dbox\}\\
&\cup&\{((s,\tphdL{\delta_L}d\delta_R),[C_{s,f}\parallel T_{\tphdL{\delta_L}d\delta_R}]_{\{r,w,L,R\}})\mid s\in\Sta_M,\,\delta_L,\delta_R\in\Dbox^{*},\,d\in\Dbox,\,\delta_L\neq\epsilon,\,\delta_L=\delta_L'f\}\\
&\cup&\{((s,\delta_Ld\tphdR{\delta_R}),[C_{s,f}\parallel T_{\delta_Ld\tphdR{\delta_R}}]_{\{r,w,L,R\}})\mid s\in\Sta_M,\,\delta_L,\delta_R\in\Dbox^{*},\,d\in\Dbox,\,\delta_R\neq\epsilon,\,\delta_R=f\delta_R'\}\\
&\cup&\{((s,\tphd{\Box}\delta_R),[C_{s,\Box}\parallel T_{\tphd{\Box}\delta_R}]_{\{r,w,L,R\}})\mid s\in\Sta_M,\,\delta_R\in\Dbox^{*}\}\\
&\cup&\{((s,\delta_L\tphd{\Box}),[C_{s,\Box}\parallel T_{\delta_L\tphd{\Box}}]_{\{r,w,L,R\}})\mid s\in\Sta_M,\,\delta_L\in\Dbox^{*}\}
\enskip.
\end{eqnarray*}

We show that $\R\subseteq\bbisimd$.

$(s,\delta_L\tphd{d}\delta_R)$ has the following transitions:
\begin{eqnarray*}
&(s,\delta_L\tphd{d}\delta_R)\step{a}(t,\tphdL{\delta_L}e\delta_R)&\mbox{if}\,(s,d,a,e,L,t)\in\step{}_{\M},\,\delta_L\neq\epsilon\\
&(s,\delta_L\tphd{d}\delta_R)\step{a}(t,\delta_Le\tphdR{\delta_R})&\mbox{if}\,(s,d,a,e,R,t)\in\step{}_{\M},\,\delta_R\neq\epsilon\\
&(s,\delta_L\tphd{d}\delta_R)\step{a}(t,\tphd{\Box}e\delta_R)&\mbox{if}\,(s,d,a,e,L,t)\in\step{}_{\M},\,\delta_L=\epsilon\\
&(s,\delta_L\tphd{d}\delta_R)\step{a}(t,\delta_Le\tphd{\Box})&\mbox{if}\,(s,d,a,e,R,t)\in\step{}_{\M},\,\delta_R=\epsilon
\enskip.
\end{eqnarray*}

They are simulated by:
\begin{eqnarray*}
&&[C_{s,d}\parallel T_{\delta_L\tphd{d}\delta_R}]_{\{r,w,L,R\}}\step{a}[w!e.L!m.\Sigma_{f\in\Dbox}r?f.C_{t,f}\parallel T_{\delta_L\tphd{d}\delta_R}]_{\{r,w,L,R\}}\\
&&\step{}^{*}[C_{t,f}\parallel T_{\tphdL{\delta_L}d\delta_R}]_{\{r,w,L,R\}},\,\mbox{if}\,(s,d,a,e,L,t)\in\step{}_{\M},\,\delta_L\neq\epsilon,\,\delta_L=\delta_L'f\\
&&[C_{s,d}\parallel T_{\delta_L\tphd{d}\delta_R}]_{\{r,w,L,R\}}\step{a}[w!e.R!m.\Sigma_{f\in\Dbox}r?f.C_{t,f}\parallel T_{\delta_L\tphd{d}\delta_R}]_{\{r,w,L,R\}}\\
&&\step{}^{*}[C_{t,f}\parallel T_{\delta_Ld\tphdR{\delta_R}}]_{\{r,w,L,R\}},\,\mbox{if}\,(s,d,a,e,R,t)\in\step{}_{\M},\,\delta_R\neq\epsilon,\,\delta_R=f\delta_R'\\
&&[C_{s,d}\parallel T_{\delta_L\tphd{d}\delta_R}]_{\{r,w,L,R\}}\step{a}[w!e.L!m.\Sigma_{f\in\Dbox}r?f.C_{t,f}\parallel T_{\delta_L\tphd{d}\delta_R}]_{\{r,w,L,R\}}\\
&&\step{}^{*}[C_{t,\Box}\parallel T_{\tphd{\Box}d\delta_R}]_{\{r,w,L,R\}},\,\mbox{if}\,(s,d,a,e,L,t)\in\step{}_{\M},\,\delta_L=\epsilon\\
&&[C_{s,d}\parallel T_{\delta_L\tphd{d}\delta_R}]_{\{r,w,L,R\}}\step{a}[w!e.R!m.\Sigma_{f\in\Dbox}r?f.C_{t,f}\parallel T_{\delta_L\tphd{d}\delta_R}]_{\{r,w,L,R\}}\\
&&\step{}^{*}[C_{t,\Box}\parallel T_{\delta_Ld\tphd{\Box}}]_{\{r,w,L,R\}},\,\mbox{if}\,(s,d,a,e,L,t)\in\step{}_{\M},\,\delta_R=\epsilon
\enskip.
\end{eqnarray*}

We apply similar analysis to other pairs in $\R$. Using the proof strategy similar to Lemma~\ref{lemma:tcpn-stack}, it is straightforward show that $\R$ is a bisimulation up to $\bbisim$. Hence, we have $\R\subset\bbisim$. Moreover, using a similar strategy in the proof showing a $\pi$-calculus is reactively Turing powerful~\cite{LY14}, we can show that $\R$ satisfies the divergence-preserving condition. For every infinite $\tau$-transition sequence in $\T(\M)$, we can find an infinite $\tau$-transition sequence in the transition system induced from $[C_{\uparrow_{\M},\Box}\parallel T]_{\{r,w,L,R\}}$.
Therefore, $\R\subset\bbisimd$.

Hence, we have $\T(\M)\bbisimd [C_{\uparrow_{\M},\Box}\parallel T]_{\{r,w,L,R\}}$.
\end{proof}}

We have the following theorem.

\begin{theorem}~\label{thm:tcpn}
\TCPN{} is reactively Turing powerful modulo $\bbisimd$.
\end{theorem} 
\section{Conclusion}\label{sec:conclusion}
In this paper we have proposed a revised  operational semantics of the sequential composition operator in the presence of intermediate termination. We established two results which is still unsolved with the standard version of the sequential composition operator. We first proved that, with the revised semantics, every context-free process corresponds to a pushdown process modulo strong bisimilarity. We also proved that, \TCPN{} is a reactively Turing powerful process calculi modulo divergence-preserving branching bisimilarity.

There are still some negative premise in the revised sequential composition operator. For instance, unguarded recursion causes problems. Consider the following process:
\begin{equation*}
P_1=P_1\bullet P_2+\one\quad
P_2=a.\one
\enskip.
\end{equation*}
According to our operational semantics, no transition is allowed from $P_1$. If we replace $\bullet$ by $\cdot$, $P_1$ would be able to do an $a$-labelled transition resulting in infinitely many distinct states. We do not have a perfect solution on dealing with transitions from processes specified with unguarded recursions yet.

Moreover, the congruence property only holds on the rooted divergence-preserving branching bisimilarity. It fails for the rooted divergence-insensitive version of  branching bisimilarity on \TCPS{}. Consider the following processes:
\begin{equation*}
P_1=\tau.\one\quad P_2=(\tau.\one)^{*}\quad Q=a.\one
\enskip.
\end{equation*}
We have $P_1\rbbisim P_2$ but not $P_1\bullet Q\rbbisim P_2\bullet Q$. Since the first process have an $a$-labelled transition followed by a $\tau$-labelled transition, but the second process only have $\tau$-labelled transitions.

Moreover, the standard semantics is designed to satisfy the axiom $(x+y)\cdot z=x\cdot z+y\cdot z$. However, it is no longer valid in the revised semantics. For instance, $(a+\one)\bullet b$ is no longer equivalent to $a\bullet b+\one\bullet b$ modulo any behavioural equivalence.
In the future work, we shall provide a sound and complete axiomatisation for the process calculus \TCPS{} with respect to strong bisimilarity as well as rooted divergence-preserving branching bisimilarity.

 Another interesting future work is to establish reactive Turing powerfulness on other process calculi with non-regular iterators based on the revised semantics of the sequential composition operator. For instance, we could consider the pushdown operator ``$\sharp$'' and the back-and-forth operator ``$\leftrightarrows$'' introduced by Bergstra and Ponse in~\cite{bergstra2001non}. They are defined by the following equations:
\begin{equation*}
\pushd{P_1}{P_2}= P_1\bullet (\pushd{P_1}{P_2})\bullet (\pushd{P_1}{P_2})+P_2\quad
\backf{P_1}{P_2}= P_1\bullet (\backf{P_1}{P_2})\bullet P_2+P_2
\enskip.
\end{equation*}
By analogy to the nesting operator, we shall also give a proper operational semantics, and then use the calculus obtained by the revised semantics to define other versions of terminating counters.

\bibliographystyle{eptcs}
\bibliography{Executability}
\conf{
\appendix
\newpage
\section{Proof in Section~\ref{sec:sequential}}\label{sec:Proof3}
\myparagraph{Proof of Theorem~\ref{thm:congruence}}
\begin{proof}
We use the following facts:
\begin{enumerate}
\item  Rooted divergence-preserving branching bisimilarity
is a rooted divergence-preserving branching
bisimulation relation; and
\item rooted divergence-preserving branching bisimilarity
is a subset of divergence-preserving branching
bisimilarity.
\end{enumerate}
We show that $\rbbisimd$ is compatible for each operator $a.,+,\bullet,\parallel$.
\begin{enumerate}
\item Suppose that $P\rbbisimd Q$, we show that $a.P\rbbisimd a.Q$. To this end, we verify that $\R=\{(a.P,a.Q)\mid P\rbbisimd Q\}\cup{\rbbisimd}$ is a rooted divergence-preserving branching bisimulation relation.

    To prove that the pair $(a.P, a.Q)$ with $P$ rooted divergence-preserving branching bisimilar to Q satisfies
condition 1 of Definition~\label{def:root}, suppose that $a.P\step{b}P'$. Then, according to the operational semantics, $b=a$
and $P'=P$. By the operational semantics, we also have that $a.Q\step{a} Q$ and, by assumption, $P$ and $Q$ are
divergence-preserving branching bisimilar.

    For the termination condition, it is trivially satisfied since both processes do not terminate. The divergence-preserving condition is also satisfied since only an $a$-labelled transition is allowed from both processes.

\item Suppose that $P_1\rbbisimd Q_1$ and $P_2\rbbisimd Q_2$, we show that $P_1+P_2\rbbisimd Q_1+Q_2$. To this end, we verify that $\R=\{(P_1+P_2,Q_1+Q_2)\mid P_1\rbbisimd Q_1,\,P_2\rbbisimd Q_2\}\cup{\rbbisimd}$ is a rooted divergence-preserving branching bisimulation relation.

    Suppose that $P_1+P_2\step{a}P'$; then we have $P_1\step{a}P'$ or $P_2\step{a}P'$. We only consider the first case.
    Since $P_1\rbbisimd Q_1$, we have $Q_1\step{a}Q'$ with $P'\bbisimd Q'$. Then we have $Q_1+Q_2\step{a}Q'$ with $P'\bbisimd Q'$. The same argument holds for the symmetrical case.

    If $P_1+P_2\downarrow$ , then we have either $P_1\downarrow$ or $P_2\downarrow$. Without loss of generality, we suppose that $P_1\downarrow$. Since $P_1\rbbisimd Q_1$, we have $Q_1\downarrow$. Therefore, $Q_1+Q_2\downarrow$.

    Moreover, we verify that the divergence preservation condition is satisfied.

    Hence, $\R$ is a rooted divergence-preserving branching bisimulation relation.

\item Suppose that $P_1\rbbisimd Q_1$ and $P_2\rbbisimd Q_2$, we show that $[P_1\parallel P_2]_{\C'}\rbbisimd [Q_1\parallel Q_2]_{\C'}$. To this end, we verify that $\R=\{([P_1\parallel P_2]_{\C'},[Q_1\parallel Q_2]_{\C'})\mid P_1\rbbisimd Q_1,\,P_2\rbbisimd Q_2\}\cup{\rbbisimd}$ is a rooted divergence-preserving branching bisimulation relation.

    We first show that $\R'=\{([P_1\parallel P_2]_{\C'},[Q_1\parallel Q_2]_{\C'})\mid P_1\bbisimd Q_1,\,P_2\bbisimd Q_2\}\cup{\bbisimd}$ is a divergence-preserving branching bisimulation.

    Suppose that $[P_1\parallel P_2]_{\C'}\step{a}P'$; then we distinguish several cases.
    \begin{enumerate}
    \item If $P_1\step{a}P_1'\,a\notin \I_{\C'}$ and $P'=[P_1'\parallel P_2]_{\C'}$, then, since $P_1\bbisimd Q_1$, we have $Q_1\step{}^{*}Q_1''\step{a}Q_1'$ with $P_1'\bbisimd Q_1'$ and $P_1\bbisimd Q_1''$. Then we have $[Q_1\parallel Q_2]_{\C'}\step{}^{*}[Q_1''\parallel Q_2]_{\C'}\step{a}[Q_1'\parallel Q_2]_{\C'}$ with $P_1\bbisimd Q_1''$, $P_1'\bbisimd Q_1'$ and $P_2\bbisimd Q_2$. Thus we have $([P_1'\parallel P_2]_{\C'},[Q_1'\parallel Q_2]_{\C'})\in\R'$ and $([P_1\parallel P_2]_{\C'},[Q_1''\parallel Q_2]_{\C'})\in\R'$.

    \item If $P_1\step{c?d}P_1',\,P_2\step{c!d}P_2'$ and $ c\in\C'$, then $[P_1\parallel P_2]_{\C'}\step{\tau}[P_1'\parallel P_2']_{\C'}$. Since $P_1\bbisimd Q_1$ and $P_2\bbisimd Q_2$, we have $Q_1\step{}^{*}Q_1''\step{c?d}Q_1',\,Q_2\step{}^{*}Q_2''\step{c!d}Q_2'$ with $P_1'\bbisimd Q_1'$ , $P\bbisimd Q_1''$, $P_2'\bbisimd Q_2'$, and $P_2\bbisimd Q_2''$. Then we have $[Q_1\parallel Q_2]_{\C'}\step{}^{*}[Q_1''\parallel Q_2'']_{\C'}\step{\tau} [Q_1'\parallel Q_2']_{\C'}$ with $P_1\bbisimd Q_1''$, $P_2\bbisimd Q_2''$, $P_1'\bbisimd Q_1'$ and $P_2'\bbisimd Q_2'$. Thus we have $([P_1\parallel P_2]_{\C'},[Q_1''\parallel Q_2'']_{\C'})\in\R'$ and $([P_1'\parallel P_2']_{\C'},[Q_1'\parallel Q_2']_{\C'})\in\R'$.
    \end{enumerate}

    If $[P_1\parallel P_2]_{\C'}\downarrow$, then we have $P_1\downarrow$ and $P_2\downarrow$. Since $P_1\bbisimd Q_1$ and $P_2\bbisimd Q_2$, we have $Q_1\step{}^{*}Q_1'\downarrow$ and $Q_2\step{}^{*}Q_2'\downarrow$ for some $Q_1'$ and $Q_2'$. Therefore, $[Q_1\parallel Q_2]_{\C'}\step{}^{*}[Q_1'\parallel Q_2']_{\C'}\downarrow$.

    Hence, we have $\R'$ is a divergence-preserving branching bisimulation relation.

    Now we show that $\R$ is a rooted divergence-preserving branching bisimulation.
    Suppose that $[P_1\parallel P_2]_{\C'}\step{a}P'$; then we distinguish several cases.
    \begin{enumerate}
    \item If $P_1\step{a}P_1'\,a\notin \I_{\C'}$ and $P'=[P_1'\parallel P_2]_{\C'}$, then, since $P_1\rbbisimd Q_1$, we have $Q_1\step{a}Q_1'$ with $P_1'\bbisimd Q_1'$. Then we have $[Q_1\parallel Q_2]_{\C'}\step{a} [Q_1'\parallel Q_2]_{\C'}$ with $P_1'\bbisimd Q_1'$ and $P_2\bbisimd Q_2$. Thus we have $[P_1'\parallel P_2]_{\C'}\bbisimd [Q_1'\parallel Q_2]_{\C'}$.

    \item If $P_1\step{c?d}P_1',\,P_2\step{c!d}P_2'$ and $ c\in\C'$, then $[P_1\parallel P_2]_{\C'}\step{\tau}[P_1'\parallel P_2']_{\C'}$. Since $P_1\rbbisimd Q_1$ and $P_2\rbbisimd Q_2$, we have $Q_1\step{c?d}Q_1',\,Q_2\step{c!d}Q_2'$ with $P_1'\bbisimd Q_1'$ and $P_2'\bbisimd Q_2'$. Then we have $[Q_1\parallel Q_2]_{\C'}\step{\tau} [Q_1'\parallel Q_2']_{\C'}$ with $P_1'\bbisimd Q_1'$ and $P_2'\bbisimd Q_2'$. Thus we have $[P_1'\parallel P_2']_{\C'}\bbisimd [Q_1'\parallel Q_2']_{\C'}$.
    \end{enumerate}

    If $[P_1\parallel P_2]_{\C'}\downarrow$, then we have $P_1\downarrow$ and $P_2\downarrow$. Since $P_1\rbbisimd Q_1$ and $P_2\rbbisimd Q_2$, we have $Q_1\downarrow$ and $Q_2\downarrow$. Therefore, $[Q_1\parallel Q_2]_{\C'}\downarrow$.

    Moreover, we verify that the divergence preservation condition is satisfied.

    Hence, we have $\R$ is a rooted divergence-preserving branching bisimulation relation.
\item  Suppose that $P_1\rbbisimd Q_1$ and $P_2\rbbisimd Q_2$, we show that $P_1\bullet P_2\rbbisimd Q_1\bullet Q_2$. To this end, we verify that $\R=\{(P_1\bullet P_2,Q_1\bullet Q_2)\mid P_1\rbbisimd Q_1,\,P_2\rbbisimd Q_2\}\cup{\rbbisimd}$ is a rooted divergence-preserving branching bisimulation relation.

We first show that $\R'=\{(P_1\bullet P_2,Q_1\bullet Q_2)\mid P_1\bbisimd Q_1,\,P_2\rbbisimd Q_2\}\cup{\bbisimd}$ is a divergence-preserving branching bisimulation relation.

 Suppose that $P_1\bullet P_2\step{a}P'$; then we distinguish several cases.
    \begin{enumerate}
    \item If $P_1\step{a} P_1'$, then $P'=P_1'\bullet P_2$. Since $P_1\bbisimd Q_1$, we have $Q_1\step{}^{*}Q_1''\step{a}Q_1'$ with $P_1'\bbisimd Q_1'$ and $P_1\bbisimd Q_1''$. Then we have $Q_1\bullet Q_2 \step{}^{*}Q_1''\bullet Q_2\step{a}Q_1'\bullet Q_2$ with $P_1\bbisimd Q_1''$, $P_1'\bbisimd Q_1'$, and $P_2\rbbisimd Q_2$. Thus, we have $(P_1'\bullet P_2, Q_1'\bullet Q_2)\in\R'$ and $(P_1\bullet P_2,Q_1''\bullet Q_2)\in\R'$.
    \item If $P_1\downarrow,\, P_2\step{a}P_2'$ and $P_1\not{\step{}}$. Since $P_1\bbisimd Q_1$ and $P_2\rbbisimd Q_2$, we have $Q_1\step{}^{*}Q_1'\downarrow$, $Q_1'\not{\step{}}$ for some $Q_1'$ with $P_1\bbisimd Q_1'$, and $Q_2\step{a}Q_2'$, with $P_2'\bbisimd Q_2'$. Then, we have $Q_1\bullet Q_2\step{}^{*}Q_1'\bullet Q_2\step{a} Q_2'$ with $P_2'\bbisimd Q_2'$ and $P_1\bbisimd Q_1'$. Thus we have $(P_2',Q_2')\in\R'$ and $(P_1\bullet P_2,Q_1'\bullet Q_2)\in\R$.
    \end{enumerate}

    If $P_1\bullet P_2\downarrow$, then we have $P_1\downarrow$ and $P_2\downarrow$. Since $P_1\bbisimd Q_1$ and $P_2\rbbisimd Q_2$, we have $Q_1\step{}^{*}Q_1'\downarrow$ for some $Q_1'$ and $Q_2\downarrow$. Therefore, $Q_1\bullet Q_2\step{}^{*}Q_1'\bullet Q_2\downarrow$.

    Moreover, we verify that the divergence preservation condition is satisfied.

    Hence, we have $\R$ is a divergence-preserving branching bisimulation relation.

Now we show that $\R$ is a rooted divergence-preserving branching bisimulation relation.

    We suppose that $P_1\bullet P_2\step{a}P'$, we distinguish several cases:
    \begin{enumerate}
    \item If $P_1\step{a} P_1'$, then $P'=P_1'\bullet P_2$. Since $P_1\rbbisimd Q_1$, we have $Q_1\step{a}Q_1'$ with $P_1'\bbisimd Q_1'$. Then we have $Q_1\bullet Q_2 \step{a}Q_1'\bullet Q_2$ with $P_1'\bbisimd Q_1'$ and $P_2\rbbisimd Q_2$. Thus, we have $P_1'\bullet P_2\bbisimd Q_1'\bullet Q_2$.
    \item If $P_1\downarrow,\, P_2\step{a}P_2'$ and $P_1\not{\step{}}$. Since $P_1\rbbisimd Q_1$ and $P_2\rbbisimd Q_2$, we have $Q_1\downarrow$, $Q_2\step{a}Q_2'$, with $P_2'\bbisimd Q_2'$, and $Q_1\not{\step{}}$. Then, we have $Q_1\bullet Q_2\step{a} Q_2'$ with $P_2'\bbisimd Q_2'$.
    \end{enumerate}

    If $P_1\bullet P_2\downarrow$, then we have $P_1\downarrow$ and $P_2\downarrow$. Since $P_1\rbbisimd Q_1$ and $P_2\rbbisimd Q_2$, we have $Q_1\downarrow$ and $Q_2\downarrow$. Therefore, $Q_1\bullet Q_2\downarrow$.

    Moreover, we verify that the divergence preservation condition is satisfied.

    Hence, we have $\R$ is a rooted divergence-preserving branching bisimulation relation.
\end{enumerate}
\end{proof} 
\section{Proof in Section~\ref{sec:cfg}}\label{sec:Proof4}
\myparagraph{Proof of Lemma~\ref{lemma:cfp-pda}}
\begin{proof}
We first define an auxiliary function $\mathit{stack}:\V^{*}\rightarrow \D^{*}$ as follows: given $\xi\in\V^{*}$, for $k=1,\ldots,\length{\xi}$, we let $X_k=\get{\xi}{k}$,
\begin{enumerate}
\item  if $X_k\notin\suffset{\xi}{k}$, then the $k$-th the element of $\stack{\xi}$ is $X^{\dagger}_{k}$,
\item otherwise, the $k$-the the element of $\stack{\xi}$ is $X_{k}$,
\end{enumerate}
Note that $\stack{X\xi}$ and $\stack{\xi}$ share the same suffix of length $\length{\xi}$.

We show that the following relation:
\begin{equation*}
\R=\{(\xi,(s_{\suffset{\xi}{0}},\mathit{stack}(\xi)))\mid \xi\in\V^{*}\}
\enskip,
\end{equation*}
is a strong bisimulation.

We rewrite $\xi$ as $X_j\xi'$, then it has the following transitions:
\begin{equation*}
X_j\xi'\step{\alpha_{ij}} \xi_{ij}\xi',\,i\in I_{X_j}
\enskip.
\end{equation*}

We need to show that they are simulated by the transitions:
\begin{equation*}
(s_{\suffset{\xi}{0}},\stack{\xi})\step{\alpha_{ij}}(s_{\suffset{\xi_{ij}\xi'}{0}},\stack{\xi_{ij}\xi'}),\,i\in I_{X_j}
\enskip.
\end{equation*}
Thus we have $(x_{ij},(s_{\suffset{\xi_{ij}\xi'}{0}},\stack{\xi_{ij}\xi'}))\in\R$.

We consider the configuration $(s_{\suffset{\xi}{0}},\stack{\xi})$, we distinguish two cases of the top symbol of the stack.
\begin{enumerate}
\item If $\get{\stack{\xi}}{1}=X^{\dagger}_{j}$, then $\M$ has the transition
\begin{equation*}
(s_{\suffset{\xi}{0}},X^{\dagger}_{j},\alpha_{ij},\delta(s_{\suffset{\xi}{0}},X^{\dagger}_{j},\xi_{ij}),s_{D(s_{\suffset{\xi}{0}},X^{\dagger}_{j},\xi_{ij})})
\enskip.
\end{equation*}
The new stack is $S=\delta(s_{\suffset{\xi}{0}},X^{\dagger}_{j},\xi_{ij})\stack{\xi'}$. We verify that $S=\stack({\xi_{ij}\xi'})$. Note that they share the same suffix $\stack{\xi'}$. We only needs to verify the first $\length{\xi_{ij}}$ elements. For the $l$-th element, we let $X_l=\get{\xi{ij}}{l}$, and we distinguish with two cases.
\begin{enumerate}
\item If $X_l\notin(\suffset{\xi}{0}/\{X_j\})\cup \suffset{\xi_{ij}}{l}$, then the $l$-th element of $S$ is $X^{\dagger}_{l}$. Since $\get{\stack{\xi}}{1}=X^{\dagger}_{j}$, from the definition of $\mathit{stack}$, we have $X_j\notin\suffset{\xi}{1}=\suffset{\xi'}{0}$. Therefore, $\suffset{\xi}{0}/\{X_j\}=\suffset{\xi'}{0}$. In this case, $X_l\notin\suffset{\xi'}{0}\cup\suffset{\xi_{ij}}{l}$. Moreover, we have $X_l\notin\suffset{\xi_{ij}\xi'}{l}$, therefore, the $l$-th element of $\stack{\xi_{ij}\xi'}$ is also $X^{\dagger}_l$.
\item Otherwise, then the $l$-th element of $S$ is $X_{l}$. By the definition of $\mathit{stack}$, we get that the $l$-th element of $\stack{\xi_{ij}\xi'}$ is also $X_l$.
\end{enumerate}
Moreover, we verify that the new state $s_{D(s_{\suffset{\xi}{0}},X^{\dagger}_{j},\xi_{ij})}=s_{\suffset{\xi_{ij}\xi'}{0}}$.
Note that we have
\begin{eqnarray*}
&&D(s_{\suffset{\xi}{0}},X^{\dagger}_{j},\xi_{ij})=(\suffset{\xi}{0}/\{X_j\})\cup\suffset{\xi_{ij}}{0}\\
&&=\suffset{\xi'}{0}\cup\suffset{\xi_{ij}}{0}=\suffset{\xi_{ij}\xi'}{0}
\enskip.
\end{eqnarray*}

Hence, we have $(s_{\suffset{\xi}{0}},\stack{\xi})\step{\alpha_{ij}}(s_{\suffset{\xi_{ij}\xi'}{0}},\stack{\xi_{ij}\xi'})$.
\item if  $\get{\stack{\xi}}{1}=X_{j}$, then $\M$ has the transition
\begin{equation*}
(s_{\suffset{\xi}{0}},X_{j},\alpha_{ij},\delta(s_{\suffset{\xi}{0}},X_{j},\xi_{ij}),s_{D(s_{\suffset{\xi}{0}},X_{j},\xi_{ij})})
\enskip,
\end{equation*}
The ne stack is $S=\delta(s_{\suffset{\xi}{0}},X_{j},\xi_{ij})\stack{\xi'}$. We verify that $S=\stack({\xi_{ij}\xi'})$. Note that they share the same suffix $\stack{\xi'}$. We only needs to verify the first $\length{\xi_{ij}}$ elements. For the $l$-th element, we let $X_l=\get{\xi{ij}}{l}$, and we distinguish with two cases.
\begin{enumerate}
\item If $X_l\notin(\suffset{\xi}{0})\cup \suffset{\xi_{ij}}{l}$, then the $l$-th element of $S$ is $X^{\dagger}_{l}$. Since $\get{\stack{\xi}}{1}=X_{j}$, from the definition of $\mathit{stack}$, we have $X_j\in\suffset{\xi}{1}=\suffset{\xi'}{0}$. Therefore, $\suffset{\xi}{0}=\suffset{\xi'}{0}$. In this case, $X_l\notin\suffset{\xi'}{0}\cup\suffset{\xi_{ij}}{l}$. Moreover, we have $X_l\notin\suffset{\xi_{ij}\xi'}{l}$, therefore, the $l$-th element of $\stack{\xi_{ij}\xi'}$ is also $X^{\dagger}_l$.
\item Otherwise, then the $l$-th element of $S$ is $X_{l}$. By the definition of $\mathit{stack}$, we get that the $l$-th element of $\stack{\xi_{ij}\xi'}$ is also $X_l$.
\end{enumerate}
Moreover, we verify that the new state $s_{D(s_{\suffset{\xi}{0}},X_{j},\xi_{ij})}=s_{\suffset{\xi_{ij}\xi'}{0}}$.
Note that we have
\begin{eqnarray*}
&&D(s_{\suffset{\xi}{0}},X_{j},\xi_{ij})=\suffset{\xi}{0}\cup\suffset{\xi_{ij}}{0}\\
&&=\suffset{\xi'}{0}\cup\suffset{\xi_{ij}}{0}=\suffset{\xi_{ij}\xi'}{0}
\enskip.
\end{eqnarray*}
Hence, we have $(s_{\suffset{\xi}{0}},\stack{\xi})\step{\alpha_{ij}}(s_{\suffset{\xi_{ij}\xi'}{0}},\stack{\xi_{ij}\xi'})$.
\end{enumerate}
By concluding the two cases, the above transitions are correct.

Using a similar analysis, we also have all the transitions from $(s_{\suffset{\xi}{0}},\stack{\xi})$ are simulated by $X_j\xi'$.

Now we consider the termination condition. $\xi\downarrow$ iff for all $X\in\suffset{\xi}{0}$, $X\downarrow$. Note that $(s_{\suffset{\xi}{0}},\stack{\xi})\downarrow$ iff for all $X\in\suffset{\xi}{0}$, $X\downarrow$. Therefore, termination condition is also verified.

Hence, we have $\T(X_0)\bisim \T(\M)$.
\end{proof} 
\section{Proof in Section~\ref{sec:Termination}}\label{sec:Proof5}
\myparagraph{Proof of Lemma~\ref{lemma:up-to}}
\begin{proof}
It is sufficient to prove that $\mathrel{\bbisim \mathrel{\circ} \mathrel{\R}}$ is a branching bisimulation, for $\bbisim$ is an equivalence relation.
Let $s_1,s_2,s_3\in\Sta$ and $s_1\bbisim s_2\mathrel{\R} s_3$.
\begin{enumerate}
    \item Suppose $s_1\step{a}s_1'$. We distinguish two cases:
    \begin{enumerate}
        \item If $a=\tau$ and $s_1\bbisim s_1'$, then $s_1'\bbisim s_1\bbisim s_2$, so $s_1'\mathrel{\bbisim \mathrel{\circ} \mathrel{\R}}s_3$. It satisfies Condition 1 of the definition of branching bisimulation.
        \item Otherwise, we have ${a\neq\tau}\vee{s_1\not\bbisim s_1'}$. Then, since $s_1 \bbisim s_2$, according to Definition~\ref{def:bbisim}, there exist $s_2''$ and $s_2'$ such that $s_2\step{}^{*}s_2''\step{a}s_2'$, $s_1\bbisim s_2''$ and $s_1'\bbisim s_2'$. Note that $s_2\bbisim s_1 \bbisim s_2''$, and it is needed to apply Condition 1 of Definition~\ref{def:up-to}. Then we have there exist $s_4''$, $s_4'$ and $s_3'$ such that $s_3\step{a}s_3'$ and $s_2''\bbisim s_4'' \mathrel{\R} s_3$ and $s_2'\bbisim s_4'\mathrel{\R} s_3'$. Since $s_1'\bbisim s_2'\bbisim s_4'$ and $s_4' \mathrel{\R} s_3' $, it follows that $s_1'\mathrel{\bbisim \mathrel{\circ} \mathrel{\R}} s_3'$. It satisfies Condition 1 of the definition of branching bisimulation.

    \end{enumerate}
    \item If $s_3\step{a}s_3'$, then according to Definition~\ref{def:up-to}, there exist $s_2''$ and $s_2'$ such that $s_2\step{}^{*}s_2''\step{a}s_2'$, $s_2''\bbisim s_2$ and $s_2'\mathrel{\bbisim \mathrel{\circ} \mathrel{\R}}s_3'$ , since $s_1\bbisim s_2\bbisim s_2''$ and $s_2''\step{a}s_2'$, by Definition~\ref{def:bbisim}, there exist $s_1''$ and $s_1'$ such that $s_1\step{}^{*}s_1''\step{(a)}s_1'$ with $s_1'' \bbisim s_2''$ and $s_1'\bbisim s_2'$. Since $s_2''\mathrel{\bbisim \mathrel{\circ}\mathrel{\R}} s_3$ and $s_2'\mathrel{\bbisim \mathrel{\circ} \mathrel{\R}}s_3'$, it follows that $s_1''\mathrel{\bbisim \mathrel{\circ} \mathrel{\R}}s_3$ and $s_1'\mathrel{\bbisim \mathrel{\circ} \mathrel{\R}}s_3'$. It satisfies the symmetry of Condition 1 of the definition of branching bisimulation.
\end{enumerate}
The termination condition is also satisfied from Definition~\ref{def:up-to}.

Therefore, a branching bisimulation up to $\bbisim$ is included in $\bbisim$.
\end{proof}

\myparagraph{Proof of Lemma~\ref{lemma:tcpn-halfcounter}}
\begin{proof}

We verify that $HC\bbisimd C_0$. Consider the following relation:
\begin{equation*}
\R_1=\{(C_0, HC)\}\cup\{(C_n,\nest{(a+\one)}{(b+\one)}\bullet (a+\one)^n\bullet (c+\one)\bullet HC)\mid n\geq 1\}\cup\{(B_n,(a+\one)^n\bullet (c+\one)\bullet HC)\mid n\in\mathbb{N}\}
\enskip.
\end{equation*}

We let $\R_2$ be the symmetrical relation of $\R_1$. We show that $\R=\R_1\cup\R_2$ is a divergence-preserving branching bisimulation as follows:

Note that $\R$ satisfies the divergence-preserving condition since there is no infinite sequence of $\tau$ transitions.
In this prove, we only illustrate the pairs in $\R_1$, since we can use the symmetrical argument for the pairs in $\R_2$. We first consider the pair $(C_0,HC)$. Note that $C_0$ has the following transitions:
\begin{eqnarray*}
&C_0\step{a} C_1,\,\mbox{and}\\
&C_0\step{b} B_0
\enskip,
\end{eqnarray*}
which are simulated by:
\begin{eqnarray*}
&HC\step{a}\nest{(a+\one)}{(b+\one)}\bullet (a+\one)\bullet (c+\one)\bullet HC,\, \mbox{and}\\
&HC\step{b}(c+\one)\bullet HC
\enskip,
\end{eqnarray*}
with $(C_1,\nest{(a+\one)}{(b+\one)}\bullet (a+\one)\bullet (c+\one)\bullet HC)\in\R$ and $(B_0,(c+\one)\bullet HC)\in\R$. Moreover, we have $C_0\downarrow$ and $HC\downarrow$.

Now we consider the pair $(C_n,\nest{(a+\one)}{(b+\one)}\bullet (a+\one)^n\bullet (c+\one)\bullet HC)$, with $n\geq 1$. Note that $C_n$ has the following transitions:
\begin{eqnarray*}
&C_n\step{a}C_{n+1},\,\mbox{and}\\
&C_n\step{b}B_{n}
\enskip,
\end{eqnarray*}
which are simulated by:
\begin{eqnarray*}
&\nest{(a+\one)}{(b+\one)}\bullet (a+\one)^n\bullet (c+\one)\bullet HC\step{a} \nest{(a+\one)}{(b+\one)}\bullet (a+\one)^{n+1}\bullet (c+\one)\bullet HC,\,\mbox{and}\\
&\nest{(a+\one)}{(b+\one)}\bullet (a+\one)^n\bullet (c+\one)\bullet HC\step{b}(a+\one)^n\bullet (c+\one)\bullet HC
\enskip,
\end{eqnarray*}
with $(C_{n+1},\nest{(a+\one)}{(b+\one)}\bullet (a+\one)^{n+1}\bullet (c+\one)\bullet HC)\in\R$ and $(B_n,(a+\one)^n\bullet (c+\one)\bullet HC)\in\R$. Moreover, we have $C_n\downarrow$ and $\nest{(a+\one)}{(b+\one)}\bullet (a+\one)^n\bullet (c+\one)\bullet HC\downarrow$.

Now we proceed to consider the pair $(B_0,(c+\one)\bullet HC)$. Note that $B_0$ has the following transition:
\begin{eqnarray*}
&B_0\step{c}C_0
\enskip,
\end{eqnarray*}
which is simulated by:
\begin{eqnarray*}
&(c+\one)\bullet HC\step{c}HC
\enskip,
\end{eqnarray*}
with $(C_0,HC)\in\R$. Moreover, we have $B_0\downarrow$ and $(c+\one)\bullet HC\downarrow$.

Next we consider the pair $(B_n,(a+\one)^n\bullet (c+\one)\bullet HC)$, with $n\geq 1$. Note that $B_n$ has the following transition:
\begin{eqnarray*}
&B_n\step{a}B_{n-1}
\enskip,
\end{eqnarray*}
which is simulated by:
\begin{eqnarray*}
&(a+\one)^n\bullet (c+\one)\bullet HC\step{a}(a+\one)^{n-1}\bullet (c+\one)\bullet HC
\enskip,
\end{eqnarray*}
with $(B_{n-1},(a+\one)^{n-1}\bullet (c+\one)\bullet HC)\in\R$. Moreover, we have $B_n\downarrow$ and $(a+\one)^n\bullet (c+\one)\bullet HC\downarrow$.

Hence, we have $C_0\bbisimd HC$.
\end{proof}

\myparagraph{Proof of Lemma~\ref{lemma:tcpn-rular}}
\begin{proof}
We consider a regular process with at finite set of action labels $\Atau$ which is given by $P_i=\sum_{j=1}^{n} \alpha_{ij}\bullet P_j +\beta_i\,(i=1,\ldots,n)$ where $\alpha_{ij}$ and $\beta_{i}$ are finite sums of actions from $\Atau$.  We let $c!0,c!1,\ldots,c!(n+1),c?0,c?1,\ldots,c?(n+1)$ be labels that are not in $\Atau$.

Consider the following process:
\begin{eqnarray*}
G_i&=&\sum_{j=1}^{n}\alpha_{ij}\bullet (c!j+\one) +\beta_i\bullet (c!0+\one)\\
M&=&\nest{\left(\sum_{j=1}^{n}(c?j+\one)\bullet G_j+(c!(n+1)+\one)\bullet(c?(n+1)+\one)\right)}{(c?0+\one)}\\
N&=&\nest{\left(\sum_{j=1}^{n+1}(c?j+\one)\bullet(c!j+\one)\right)}{((c?0+\one)\bullet(c!0+\one))}
\end{eqnarray*}

Note that $\bullet$ is associative and we suppose that $\bullet$ binds stronger than $+$.
We verify that $P_i\bbisimd [G_i\bullet M\parallel N]_{\{c\}}$.
We let $Q=\left(\sum_{j=1}^{n}(c?j+\one)\bullet G_j+(c!(n+1)+\one)\bullet (c?(n+1)+\one)\right)$ and $O=\left(\sum_{j=1}^{n+1}(c?j+\one)\bullet(c!j+\one)\right)$. We let
\begin{eqnarray*}
\R_1&=&\{(P_i, [G_i\bullet M\bullet Q^{k}\parallel N\bullet O^{k}]_{\{c\}})\mid k\in\mathbb{N},\,i=1,\ldots,n\}\\
&\cup&\{(P_i,[(c!i+\one)\bullet M\bullet Q^{k}\parallel N\bullet O^{k}]_{\{c\}})\mid k\in\mathbb{N},\,i=1,\ldots,n\}\\
&\cup&\{(P_i,[M\bullet Q^{k}\parallel (c!i+\one)\bullet N\bullet O^{k+1}]_{\{c\}})\mid k\in\mathbb{N},\,i=1,\ldots,n\}\\
&\cup&\{(\one,[(c!0+\one)\bullet M\bullet Q^{k}\parallel N\bullet O^{k}]_{\{c\}})\mid k\in\mathbb{N}\}\\
&\cup&\{(\one,[M\bullet Q^{k}\parallel (c!0+\one)\bullet O^{k}]_{\{c\}})\mid k\in\mathbb{N}\}\\
&\cup&\{(\one,[Q^{k}\parallel O^{k}]_{\{c\}})\mid k\in\mathbb{N}\}\\
&\cup&\{(\one,[(c?(n+1)+\one)\bullet Q^{k}\parallel (c!(n+1)+\one)\bullet O^{k}]_{\{c\}})\mid k\in\mathbb{N}\}
\enskip;
\end{eqnarray*}
and we let $\R_2$ be the symmetrical relation of $\R_1$. We show that $\R=\R_1\cup\R_2$ is a divergence-preserving branching bisimulation. We shall only verify the pairs in $\R_1$ in this proof since $\R$ is symmetrical.

For the set of pairs $\{(P_i, [G_i\bullet M\bullet Q^{k}\parallel N\bullet O^{k}]_{\{c\}})\mid k\in\mathbb{N},\,i=1,\ldots,n\}$, note that $P_i$ has the following transitions: $P_i\step{a}P_j$ if $a$ is a summand of $\alpha_{ij}$, or $P_i\step{a}\one$ if $a$ is a summand of $\beta_j$.

The first transition is simulated by the following transitions:
\begin{eqnarray*}
&&[G_i\bullet M\bullet Q^{k}\parallel N\bullet O^{k}]_{\{c\}}\step{a}[(c!j+\one)\bullet M\bullet Q^{k}\parallel N\bullet O^{k}]_{\{c\}}\\
&&\step{\tau}[M\bullet Q^{k}\parallel (c!j+\one) \bullet N\bullet O^{k+1}]_{\{c\}}\\
&&\step{\tau} [G_j\bullet M\bullet Q^{k+1}\parallel N\bullet O^{k+1}]_{\{c\}}
\enskip.
\end{eqnarray*}

If $k\geq 1$, then the second transition is simulated by the following transitions:
\begin{eqnarray*}
&&[G_i\bullet M\bullet Q^{k}\parallel N\bullet O^{k}]_{\{c\}}\step{a}[(c!0+\one)\bullet M\bullet Q^{k}\parallel N\bullet O^{k}]_{\{c\}}\\
&&\step{\tau}[M\bullet Q^{k}\parallel (c!0+\one)\bullet O^{k}]_{\{c\}}\step{\tau}[Q^{k}\parallel O^{k}]_{\{c\}}\\
&&\step{\tau}[(c?(n+1)+\one)\bullet Q^{k-1}\parallel (c!(n+1)+\one)\bullet O^{k-1}]_{\{c\}}\step{\tau}[Q^{k-1}\parallel O^{k-1}]_{\{c\}}\\
&&\step{}^{*}\one
\enskip;
\end{eqnarray*}

otherwise, if $k=0$, then the second transition are simulated by:

\begin{eqnarray*}
&&[G_i\bullet M\parallel N]_{\{c\}}\step{a}[(c!0+\one)\bullet M\parallel N]_{\{c\}}\\
&&\step{\tau}[M\parallel (c!0+\one)]_{\{c\}}\step{\tau}\one
\enskip.
\end{eqnarray*}

We have that that $(P_j,[(c!j+\one)\bullet M\bullet Q^{k}\parallel N\bullet O^{k}]_{\{c\}})\in\R$, $(P_j,[M\bullet Q^{k}\parallel (c!j+\one) \bullet N\bullet O^{k+1}]_{\{c\}})\in\R$, $(P_j,[G_j\bullet M\bullet Q^{k+1}\parallel N\bullet O^{k+1}]_{\{c\}})\in\R$, $(\one,[(c!0+\one)\bullet M\bullet Q^{k}\parallel N\bullet O^{k}]_{\{c\}})\in\R$, $(\one,[M\bullet Q^{k}\parallel (c!0+\one)\bullet O^{k}]_{\{c\}})\in\R$, $(\one,[Q^{k}\parallel O^{k}]_{\{c\}})$, $(\one,[(c?(n+1)+\one)\bullet Q^{k}\parallel (c!(n+1)+\one)\bullet O^{k}]_{\{c\}})\in\R$ and $(\one,\one)\in\R$ for all $k\in\mathbb{N}$ and $i,j=1,\ldots, n$.

One can easily verify that all the other pairs satisfy the condition of branching bisimulation. The relation $\R$ also satisfies the divergence-preserving condition since no infinite $\tau$-transition sequence is allowed from any process defined in $\R$.

Therefore, we get a finite specification of every regular process in \TCPN{} modulo $\bbisimd$.
\end{proof}

\myparagraph{Proof of Lemma~\ref{lemma:tcpn-stack}}
\begin{proof}
We define some auxiliary process:
\begin{eqnarray*}
P_j(0)&=&(\nest{(a_j!a+\one)}{(b_j!b+\one)}\bullet(c_j!c+\one))^{*}\,(j=1,2)\\
P_j(n)&=&\nest{(a_j!a+\one)}{(b_j!b+\one)}\bullet (a_j!a+\one)^{n}\bullet(c_j!c+\one))\bullet P_j,\,(j=1,2;\,n=1,2,\ldots)\\
Q_j(n)&=&(a_j!a+\one)^{n}\bullet(c_j!c+\one))\bullet P_j,\,(j=1,2;\,n\in\mathbb{N})
\enskip.
\end{eqnarray*}
$P_0$ and $P_1$ behave as two half counters.

We let $\R_1=\{(S_{\epsilon},S)\}\cup\{(S_{d_j\delta},[X_{j}\bullet X_{\epsilon}\parallel Q_1(m)\parallel P_2(0)]_{\{a_1,a_2,b_1,b_2,c_1,c_2\}})\mid j=\encode{d_j},m=\encode{d_j\delta},d\in\Dbox,\delta\in \Dbox^{*}\}$. We let $\R_2$ be the symmetrical relation of $\R_1$. We verify that $\R=\R_1\cup\R_2\cup\bbisimd$ is a divergence-preserving branching bisimulation relation.

Note that $S_{\epsilon}$ has the following transitions:
\begin{eqnarray*}
&&S_{\epsilon}\step{\mathit{push}?d_j}S_{d_j}\mbox{ for all }j=1,2,\ldots,N,\mbox{ and}\\
&&S_{\epsilon}\step{\mathit{pop}!\Box}S_{\epsilon}
\enskip.
\end{eqnarray*}

They are simulated by the following transitions:
\begin{eqnarray*}
&&S\step{\mathit{push}?d_j}[(a_1?a+\one)^{j}\bullet(b_1+\one)\bullet X_j\bullet X_{\epsilon}\parallel P_1(0)\parallel P_2(0)]_{\{a_1,a_2,b_1,b_2,c_1,c_2\}}\\
&&\step{}^{*}[(b_1+\one)\bullet X_j\bullet X_{\epsilon}\parallel P_1(j)\parallel P_2(0)]_{\{a_1,a_2,b_1,b_2,c_1,c_2\}}\\
&&\step{}^{*}[X_j\bullet X_{\epsilon}\parallel Q_1(j)\parallel P_2(0)]_{\{a_1,a_2,b_1,b_2,c_1,c_2\}}\mbox{ for all }j=1,2,\ldots,N,\mbox{ and}\\
&&S\step{\mathit{pop}!\Box}S
\enskip.
\end{eqnarray*}
We only consider the first case, since the second transition is trivial. We have $(S_{d_j},[X_j\bullet X_{\epsilon}\parallel Q_1(j)\parallel P_2(0)]_{\{a_1,a_2,b_1,b_2,c_1,c_2\}})\in\R$. We denote the sequence of transitions $[(a_1?a+\one)^{j}\bullet(b_1+\one)\bullet X_j\bullet X_{\epsilon}\parallel P_1(0)\parallel P_2(0)]_{\{a_1,a_2,b_1,b_2,c_1,c_2\}}\step{}^{*}[X_j\bullet X_{\epsilon}\parallel Q_1(j)\parallel P_2(0)]_{\{a_1,a_2,b_1,b_2,c_1,c_2\}})\in\R$ by $s_0\step{}^{*}s_m$. It is obvious that $s_0\bbisimd\ldots s_m$. Therefore, $S\step{\mathit{push}?d_j}s_0$, and $s_0\bbisimd s_m$ with $(S_{d_j},s_m)\in\R$.

Note that $S_{d_j\delta}$ has the following transitions:
\begin{eqnarray*}
&&S_{d_j\delta}\step{\mathit{push}?d_k}S_{d_kd_j\delta}\mbox{ for all }k=1,2,\ldots,N,\mbox{ and}\\
&&S_{d_j\delta}\step{\mathit{pop}!d_j}S_{d_k\delta'}\mbox{, where }d_k\delta'=\delta
\enskip.
\end{eqnarray*}

They are simulated by the following transitions:
\begin{eqnarray*}
&&[X_{j}\bullet X_{\epsilon}\parallel Q_1(\encode{d_j\delta})\parallel P_2(0)]_{\{a_1,a_2,b_1,b_2,c_1,c_2\}}\step{\mathit{push}?d_k}[\mathit{Push_k}\bullet X_{\epsilon}\parallel Q_1(\encode{d_j\delta})\parallel P_2(0)]_{\{a_1,a_2,b_1,b_2,c_1,c_2\}}\\
&&\step{}^{*}[(a_1?a+\one)^k\bullet\mathit{NShift2to1}\bullet X_{k}\bullet X_{\epsilon}\parallel P_1(0)\parallel Q_2(\encode{d_j\delta})]_{\{a_1,a_2,b_1,b_2,c_1,c_2\}}\\
&&\step{}^{*}[\mathit{NShift2to1}\bullet X_{k}\bullet X_{\epsilon}\parallel P_1(\encode{d_k})\parallel Q_2(\encode{d_j\delta})]_{\{a_1,a_2,b_1,b_2,c_1,c_2\}}\\
&&\step{}^{*}[X_{k}\bullet X_{\epsilon}\parallel Q_1(\encode{d_kd_j\delta})\parallel P_2(0)]_{\{a_1,a_2,b_1,b_2,c_1,c_2\}}\mbox{ for all }d_j,d_k\in\Dbox,\,\delta\in\Dbox^{*}\mbox{ and}\\
&&[X_{j}\bullet X_{\epsilon}\parallel Q_1(\encode{d_j\delta})\parallel P_2(0)]_{\{a_1,a_2,b_1,b_2,c_1,c_2\}}\step{\mathit{pop}!d_j}[\mathit{Pop_j}\bullet X_{\epsilon}\parallel Q_1(\encode{d_j\delta})\parallel P_2(0)]_{\{a_1,a_2,b_1,b_2,c_1,c_2\}}\\
&&\step{}^{*}[\mathit{1/NShift1to2}\bullet\mathit{Test_{\emptyset}}\bullet X_{\epsilon}\parallel Q_1(\encode{d_j\delta}-k)\parallel P_2(0)]_{\{a_1,a_2,b_1,b_2,c_1,c_2\}}\\
&&\step{}^{*}[\mathit{Test_{\emptyset}}\bullet X_{\epsilon}\parallel P_1(0)\parallel Q_2(\encode{\delta})]_{\{a_1,a_2,b_1,b_2,c_1,c_2\}}\\
&&\step{}^{*}[X_k\bullet X_{\epsilon}\parallel Q_1(\encode{d_k\delta'}\parallel P_2(0)]_{\{a_1,a_2,b_1,b_2,c_1,c_2\}}\mbox{ for all }d_j\in\Dbox,\,\delta\in\Dbox^{*}\mbox{ and}\,\delta=d_k\delta'
\enskip.
\end{eqnarray*}

We have $(S_{d_kd_j\delta},[X_{k}\bullet X_{\epsilon}\parallel Q_1(\encode{d_kd_j\delta})\parallel P_2(0)]_{\{a_1,a_2,b_1,b_2,c_1,c_2\}})\in\R$ and $(S_{d_k\delta'},[X_k\bullet X_{\epsilon}\parallel Q_1(\encode{d_k\delta'}\parallel P_2(0)]_{\{a_1,a_2,b_1,b_2,c_1,c_2\}})\in\R$. By using a similar analysis with the previous case, we conclude that $\R$ is a bisimulation up to $\bbisim$. By Lemma~\ref{lemma:up-to}, we have $\R\subseteq\bbisim$. Moreover, there is no infinite $\tau$-transition sequence from any process defined above. Therefore, $\R\subseteq\bbisimd$.

Hence, we have $S_{\epsilon}\bbisimd S$.
\end{proof}

\myparagraph{Proof of Lemma~\ref{lemma:tcpn-tape}}
\begin{proof}
We define the following auxiliary processes:
\begin{eqnarray*}
S_1(\delta)&=&[X_{1,k}\parallel Q_1(\encode{\delta})\parallel P_2(0)]_{\{a_1,a_2,b_1,b_2,c_1,c_2\}}\\
S_2(\delta)&=&[X_{2,k}\parallel Q_1(\encode{\delta})\parallel P_2(0)]_{\{a_1,a_2,b_1,b_2,c_1,c_2\}},\,\mbox{where}\,\delta=d_k\delta'
\enskip.
\end{eqnarray*}
$X_{1,k}$ and $X_{2,k}$ is obtained by renaming $\mathit{push}$ and $\mathit{pop}$ in $X_k$ to $\mathit{push_1,\,pop_1,\,push_2}$ and $\mathit{pop_2}$ respectively. We use $\overline{\delta}$ to denote the reverse sequence of $\delta$.

We verify that
\begin{equation*}
\R=\{(T_{\delta_L\tphd{d}\delta_R},[T_{d}\parallel S_1(\overline{\delta_L})\parallel S_2(\delta_R)]_{\{\mathit{push_1,pop_1,push_2,pop_2}\}})\mid d\in\Dbox,\,\delta_L,\delta_R\in\Dbox^{*}\}\subseteq\bbisimd
\enskip.
\end{equation*}

$T_{\delta_L\tphd{d}\delta_R}$ has the following transitions:
\begin{eqnarray*}
&&T_{\delta_L\tphd{d}\delta_R}\step{r!d}T_{\delta_L\tphd{d}\delta_R}\\
&&T_{\delta_L\tphd{d}\delta_R}\step{w?e}T_{\delta_L\tphd{e}\delta_R}\,\mbox{for all}\,e\in\Dbox\\
&&T_{\delta_L\tphd{d}\delta_R}\step{L?m}T_{\tphdL{\delta_L}d\delta_R}\,\mbox{if}\,\delta_L\neq\epsilon\\
&&T_{\delta_L\tphd{d}\delta_R}\step{R?m}T_{\delta_Ld\tphdR{\delta_R}}\,\mbox{if}\,\delta_R\neq\epsilon\\
&&T_{\delta_L\tphd{d}\delta_R}\step{L?m}T_{\epsilon\tphd{\Box}d\delta_R}\,\mbox{if}\,\delta_L=\epsilon\,\mbox{and}\\
&&T_{\delta_L\tphd{d}\delta_R}\step{R?m}T_{\delta_Ld\tphd{\Box}\epsilon}\,\mbox{if}\,\delta_R=\epsilon
\enskip.
\end{eqnarray*}

They are simulated by the following transitions:
\begin{eqnarray*}
&&[T_{d}\parallel S_1(\overline{\delta_L})\parallel S_2(\delta_R)]_{\{\mathit{push_1,pop_1,push_2,pop_2}\}}\step{r!d}[T_{d}\parallel S_1(\overline{\delta_L})\parallel S_2(\delta_R)]_{\{\mathit{push_1,pop_1,push_2,pop_2}\}}\\
&&[T_{d}\parallel S_1(\overline{\delta_L})\parallel S_2(\delta_R)]_{\{\mathit{push_1,pop_1,push_2,pop_2}\}}\step{e?d}[T_{e}\parallel S_1(\overline{\delta_L})\parallel S_2(\delta_R)]_{\{\mathit{push_1,pop_1,push_2,pop_2}\}}\,\mbox{for all}\,e\in\Dbox\\
&&[T_{d}\parallel S_1(\overline{\delta_L})\parallel S_2(\delta_R)]_{\{\mathit{push_1,pop_1,push_2,pop_2}\}}\step{L?m}[\mathit{Left_d}\parallel S_1(\overline{\delta_L})\parallel S_2(\delta_R)]_{\{\mathit{push_1,pop_1,push_2,pop_2}\}}\\
&&\step{}^{*}[T_{e}\parallel S_1(\overline{\delta_L'})\parallel S_2(d\delta_R)]_{\{\mathit{push_1,pop_1,push_2,pop_2}\}},\,\delta_L=\delta_L'e,\,\mbox{if}\,\delta_L\neq\epsilon\\
&&[T_{d}\parallel S_1(\overline{\delta_L})\parallel S_2(\delta_R)]_{\{\mathit{push_1,pop_1,push_2,pop_2}\}}\step{R?m}[\mathit{Right_d}\parallel S_1(\overline{\delta_L})\parallel S_2(\delta_R)]_{\{\mathit{push_1,pop_1,push_2,pop_2}\}}\\
&&\step{}^{*}[T_{e}\parallel S_1(\overline{\delta_Ld})\parallel S_2(\delta_R')]_{\{\mathit{push_1,pop_1,push_2,pop_2}\}},\,\delta_R=e\delta_R,\,\mbox{if}\,\delta_R\neq\epsilon\\
&&[T_{d}\parallel S_1(\overline{\delta_L})\parallel S_2(\delta_R)]_{\{\mathit{push_1,pop_1,push_2,pop_2}\}}\step{L?m}[\mathit{Left_d}\parallel S_1(\overline{\delta_L})\parallel S_2(\delta_R)]_{\{\mathit{push_1,pop_1,push_2,pop_2}\}}\\
&&\step{}^{*}[T_{\Box}\parallel S_1(\epsilon)\parallel S_2(d\delta_R)]_{\{\mathit{push_1,pop_1,push_2,pop_2}\}},\,\mbox{if}\,\delta_L=\epsilon\\
&&[T_{d}\parallel S_1(\overline{\delta_L})\parallel S_2(\delta_R)]_{\{\mathit{push_1,pop_1,push_2,pop_2}\}}\step{R?m}[\mathit{Right_d}\parallel S_1(\overline{\delta_L})\parallel S_2(\delta_R)]_{\{\mathit{push_1,pop_1,push_2,pop_2}\}}\\
&&\step{}^{*}[T_{\Box}\parallel S_1(\overline{\delta_Ld})\parallel S_2(\epsilon)]_{\{\mathit{push_1,pop_1,push_2,pop_2}\}},\,\mbox{if}\,\delta_R=\epsilon
\enskip.
\end{eqnarray*}

We have
\begin{eqnarray*}
&&(T_{\delta_L\tphd{d}\delta_R},[T_{d}\parallel S_1(\overline{\delta_L})\parallel S_2(\delta_R)]_{\{\mathit{push_1,pop_1,push_2,pop_2}\}})\in\R,\\ &&(T_{\delta_L\tphd{e}\delta_R},[T_{e}\parallel S_1(\overline{\delta_L})\parallel S_2(\delta_R)]_{\{\mathit{push_1,pop_1,push_2,pop_2}\}})\in\R,\\ &&(T_{\tphdL{\delta_L}d\delta_R},[T_{e}\parallel S_1(\overline{\delta_L'})\parallel S_2(d\delta_R)]_{\{\mathit{push_1,pop_1,push_2,pop_2}\}})\in\R,\\ &&(T_{\delta_Ld\tphdR{\delta_R}},[T_{e}\parallel S_1(\overline{\delta_Ld})\parallel S_2(\delta_R')]_{\{\mathit{push_1,pop_1,push_2,pop_2}\}})\in\R,\\
&&(T_{\epsilon\tphd{\Box}d\delta_R},[T_{\Box}\parallel S_1(\epsilon)\parallel S_2(d\delta_R)]_{\{\mathit{push_1,pop_1,push_2,pop_2}\}})\in\R,\,\mbox{and}\\
&&(T_{\delta_Ld\tphd{\Box}\epsilon},[T_{\Box}\parallel S_1(\overline{\delta_Ld})\parallel S_2(\epsilon)]_{\{\mathit{push_1,pop_1,push_2,pop_2}\}})\in\R
\enskip.
\end{eqnarray*}

By an analysis similar from Lemma~\ref{lemma:tcpn-stack}, we have $\R$ is a bisimulation up to $\bbisim$. Therefore, $\R\subset\bbisim$. Moreover, there is no infinite $\tau$-transition sequence from the processes defined above. Therefore, $\R\subseteq\bbisimd$.

Hence, we have $T_{\tphd{\Box}}\bbisimd T$.
\end{proof}

\myparagraph{Proof of Lemma~\ref{lemma:tcpn-control}}
\begin{proof}
By the proof of Theorem~\ref{thm:congruence}, $\bbisimd$ is compatible with parallel composition. Therefore, it is enough to show that $\T(\M)\bbisimd [C_{\uparrow,\Box}\parallel T_{\tphd{\Box}}]_{\{r,w,L,R\}}$.

We define a binary relation $\R$ by:

\begin{eqnarray*}
\R&=&\{((s,\delta_L\tphd{d}\delta_R),[C_{s,d}\parallel T_{\delta_L\tphd{d}\delta_R}]_{\{r,w,L,R\}})\mid s\in\Sta_M,\,\delta_L,\delta_R\in\Dbox^{*},\,d\in\Dbox\}\\
&\cup&\{((s,\tphdL{\delta_L}d\delta_R),[C_{s,f}\parallel T_{\tphdL{\delta_L}d\delta_R}]_{\{r,w,L,R\}})\mid s\in\Sta_M,\,\delta_L,\delta_R\in\Dbox^{*},\,d\in\Dbox,\,\delta_L\neq\epsilon,\,\delta_L=\delta_L'f\}\\
&\cup&\{((s,\delta_Ld\tphdR{\delta_R}),[C_{s,f}\parallel T_{\delta_Ld\tphdR{\delta_R}}]_{\{r,w,L,R\}})\mid s\in\Sta_M,\,\delta_L,\delta_R\in\Dbox^{*},\,d\in\Dbox,\,\delta_R\neq\epsilon,\,\delta_R=f\delta_R'\}\\
&\cup&\{((s,\tphd{\Box}\delta_R),[C_{s,\Box}\parallel T_{\tphd{\Box}\delta_R}]_{\{r,w,L,R\}})\mid s\in\Sta_M,\,\delta_R\in\Dbox^{*}\}\\
&\cup&\{((s,\delta_L\tphd{\Box}),[C_{s,\Box}\parallel T_{\delta_L\tphd{\Box}}]_{\{r,w,L,R\}})\mid s\in\Sta_M,\,\delta_L\in\Dbox^{*}\}
\enskip.
\end{eqnarray*}

We show that $\R\subseteq\bbisimd$.

$(s,\delta_L\tphd{d}\delta_R)$ has the following transitions:
\begin{eqnarray*}
&(s,\delta_L\tphd{d}\delta_R)\step{a}(t,\tphdL{\delta_L}e\delta_R)&\mbox{if}\,(s,d,a,e,L,t)\in\step{}_{\M},\,\delta_L\neq\epsilon\\
&(s,\delta_L\tphd{d}\delta_R)\step{a}(t,\delta_Le\tphdR{\delta_R})&\mbox{if}\,(s,d,a,e,R,t)\in\step{}_{\M},\,\delta_R\neq\epsilon\\
&(s,\delta_L\tphd{d}\delta_R)\step{a}(t,\tphd{\Box}e\delta_R)&\mbox{if}\,(s,d,a,e,L,t)\in\step{}_{\M},\,\delta_L=\epsilon\\
&(s,\delta_L\tphd{d}\delta_R)\step{a}(t,\delta_Le\tphd{\Box})&\mbox{if}\,(s,d,a,e,R,t)\in\step{}_{\M},\,\delta_R=\epsilon
\enskip.
\end{eqnarray*}

They are simulated by:
\begin{eqnarray*}
&&[C_{s,d}\parallel T_{\delta_L\tphd{d}\delta_R}]_{\{r,w,L,R\}}\step{a}[w!e.L!m.\Sigma_{f\in\Dbox}r?f.C_{t,f}\parallel T_{\delta_L\tphd{d}\delta_R}]_{\{r,w,L,R\}}\\
&&\step{}^{*}[C_{t,f}\parallel T_{\tphdL{\delta_L}d\delta_R}]_{\{r,w,L,R\}},\,\mbox{if}\,(s,d,a,e,L,t)\in\step{}_{\M},\,\delta_L\neq\epsilon,\,\delta_L=\delta_L'f\\
&&[C_{s,d}\parallel T_{\delta_L\tphd{d}\delta_R}]_{\{r,w,L,R\}}\step{a}[w!e.R!m.\Sigma_{f\in\Dbox}r?f.C_{t,f}\parallel T_{\delta_L\tphd{d}\delta_R}]_{\{r,w,L,R\}}\\
&&\step{}^{*}[C_{t,f}\parallel T_{\delta_Ld\tphdR{\delta_R}}]_{\{r,w,L,R\}},\,\mbox{if}\,(s,d,a,e,R,t)\in\step{}_{\M},\,\delta_R\neq\epsilon,\,\delta_R=f\delta_R'\\
&&[C_{s,d}\parallel T_{\delta_L\tphd{d}\delta_R}]_{\{r,w,L,R\}}\step{a}[w!e.L!m.\Sigma_{f\in\Dbox}r?f.C_{t,f}\parallel T_{\delta_L\tphd{d}\delta_R}]_{\{r,w,L,R\}}\\
&&\step{}^{*}[C_{t,\Box}\parallel T_{\tphd{\Box}d\delta_R}]_{\{r,w,L,R\}},\,\mbox{if}\,(s,d,a,e,L,t)\in\step{}_{\M},\,\delta_L=\epsilon\\
&&[C_{s,d}\parallel T_{\delta_L\tphd{d}\delta_R}]_{\{r,w,L,R\}}\step{a}[w!e.R!m.\Sigma_{f\in\Dbox}r?f.C_{t,f}\parallel T_{\delta_L\tphd{d}\delta_R}]_{\{r,w,L,R\}}\\
&&\step{}^{*}[C_{t,\Box}\parallel T_{\delta_Ld\tphd{\Box}}]_{\{r,w,L,R\}},\,\mbox{if}\,(s,d,a,e,L,t)\in\step{}_{\M},\,\delta_R=\epsilon
\enskip.
\end{eqnarray*}

We apply similar analysis to other pairs in $\R$. Using the proof strategy similar to Lemma~\ref{lemma:tcpn-stack}, it is straightforward show that $\R$ is a bisimulation up to $\bbisim$. Hence, we have $\R\subset\bbisim$. Moreover, using a similar strategy in the proof showing a $\pi$-calculus is reactively Turing powerful~\cite{LY15}, we can show that $\R$ satisfies the divergence-preserving condition. For every infinite $\tau$-transition sequence in $\T(\M)$, we can find an infinite $\tau$-transition sequence in the transition system induced from $[C_{\uparrow_{\M},\Box}\parallel T]_{\{r,w,L,R\}}$.
Therefore, $\R\subset\bbisimd$.

Hence, we have $\T(\M)\bbisimd [C_{\uparrow_{\M},\Box}\parallel T]_{\{r,w,L,R\}}$.
\end{proof} 
\arx{
\section{Executability without Termination}~\label{sec:NoTermination}

\todo{We need to discuss whether this section is necessary here. We do not need the new version of sequential composition here. Moreover, the results in this section is not a big improvement of the previous paper of Bergstra and Ponse.}
We consider the following variations of \TCP, namely \TCPN, \TCPP, \TCPB, \TCPR.

\todo{Introduce these calculi}

In this section, we show that from the work of Bergstra and Ponse~\cite{bergstra2001non}, we can derive the result that \TCPN, \TCPP and \TCPB are reactive Turing powerful.

We have the following statements:

\begin{enumerate}
\item If each regular process and a counter can be specified in a process calculus, then it is reactive Turing powerful (without termination)

\item  A stack can be specified by a regular process in parallel with two counters.

\item A reactive Turing machine can be specified by a regular process in parallel with two stacks.

\end{enumerate}

Hence, we can show that \TCPN, \TCPP and \TCPB are reactive Turing powerful by showing that they satisfy the first statement above.

Specifications:\\
Counter
\begin{eqnarray*}
C_0 &=& \mathit{a}.C_1+\mathit{c}.C_0\\
C_n &=& \mathit{a}.C_{n+1}+\mathit{b}.C_{n-1}\,(n\geq 1)
\end{eqnarray*}

Half-Counter
\begin{eqnarray*}
C_n &=& \mathit{a}.C_{n+1}+\mathit{b}.B_{n}\,(n\in\mathbb{N})\\
B_n &=& \mathit{a}.B_{n-1}\,(n\geq 1)\\
B_0 &=& \mathit{c}.C_0
\end{eqnarray*}

Bfi-Counter
\begin{eqnarray*}
C_n &=& \mathit{a}.C_{n+1}+\mathit{b}.B_{n}\,(n\in\mathbb{N})\\
B_n &=& \mathit{b}.B_{n-1}\,(n\geq 1)\\
B_0 &=& \mathit{c}.C_0
\end{eqnarray*}

Implementations:\\
Counter

\begin{equation*}
(a(\pushd{a}{b})+c)^{*}
\end{equation*}

Half-counter
\begin{equation*}
((\nest{a}{b})c)^{*}
\end{equation*}

Bfi-counter
\begin{equation*}
((\backf{a}{b})c)^{*}
\end{equation*}

We conjecture that \TCPR is not reactive Turing powerful.

We first consider a variation of \CCS, which is \CCSR (\CCS with replication and excluding recursive definition). It is known as not Turing powerful. As a consequence, it is not reactive Turing powerful.

\TCPR extends \CCSR by adding an operator of sequential composition. We conjecture that such extension does not make the calculus Turing powerful.

We need to show that following statements:

\begin{enumerate}
\item A counter cannot be specified in \TCPR. (conjecture)

\item The transition system of a counter is executable.

\item \TCPR is not reactive Turing powerful.
\end{enumerate}

It seems that the power of scope extrusion on local names is a key point of this question. Using a $\pi$-calculus-like schema of name communication, we can establish a counter or even a reactive Turing machine.
\todo{add reference}

We might find some idea from the following consequences:

Queue is not definable by a finite guarded recursive specification in concrete process algebra without renaming; and

Queue is definable by a finite guarded recursive specification in concrete process algebra with renaming operators.

We shall prove the following theorems.

\begin{theorem}
\begin{enumerate}
\item For every executable transition system $T$, there exits a \TCPN, \TCPP or \TCPB process $P$, such that $\T(P)\bbisimd T$.

\item There exists an executable transition system $T$, such that there does not exist a \TCPR process $P$, such that $\T(P)\bbisim T$.
\end{enumerate}
\end{theorem}

\section{Axiomatization}
\todo{This part does not belong to this paper.}
In order to get a sound and ground-complete axiomatization for \TCPS{}, we introduce an auxiliary unary operator $\mathit{NT}$. Now we consider the process calculus $\TCPS+\mathit{NT}$.

\begin{eqnarray*}
x+y=y+x\\
(x+y)+z=x+(y+z) \\
x+x=x\\
x\bullet(y\bullet z)=(x\bullet y)\bullet z \\
a.x\bullet y=a.(x\bullet y)\\
\one\bullet x=x\\
x\bullet \one=x\\
\nil\bullet x=\nil\\
(a.x+b.y+z)\bullet w=a.x\bullet w+(b.y+z)\bullet w\\
(x+\one)\bullet(y+\one)=\one+x\bullet(y+\one)\\
(x+\one)\bullet\nil= x\bullet 0\\
(x+\one)\bullet a.y=x\bullet a.y\\
(x+\one)\bullet(a.y+b/z)=x\bullet(a.y+b.z)\\
(x+\one)\bullet\NT{y}=x\bullet\NT{y}\\
\NT{\one}=\nil\\
\NT{\nil}=\nil\\
\NT{x+y}=\NT{x}+\NT{y}\\
\NT{a.x}=a.x\\
\end{eqnarray*}
} 
}
\end{document}